\documentclass[twocolumn]{aastex631}

\usepackage{appendix}

\usepackage{xcolor}%

\begin{document}

\title{The PANORAMIC Survey: Pure Parallel Wide Area Legacy Imaging with JWST/NIRCam}

\author[0000-0003-2919-7495]{Christina C.\ Williams}
\affiliation{NSF National Optical-Infrared Astronomy Research Laboratory, 950 North Cherry Avenue, Tucson, AZ 85719, USA}
\affiliation{Steward Observatory, University of Arizona, 933 North Cherry Avenue, Tucson, AZ 85721, USA}

\author[0000-0001-5851-6649]{Pascal A. Oesch}
\affiliation{Department of Astronomy, University of Geneva, Chemin Pegasi 51, 1290 Versoix, Switzerland}
\affiliation{Cosmic Dawn Center (DAWN), Copenhagen, Denmark}
\affiliation{Niels Bohr Institute, University of Copenhagen, Jagtvej 128, Copenhagen, Denmark}

\author[0000-0001-8928-4465]{Andrea Weibel}
\affiliation{Department of Astronomy, University of Geneva, Chemin Pegasi 51, 1290 Versoix, Switzerland}

\author[0000-0003-2680-005X]{Gabriel Brammer}
\affiliation{Cosmic Dawn Center (DAWN), Copenhagen, Denmark}
\affiliation{Niels Bohr Institute, University of Copenhagen, Jagtvej 128, Copenhagen, Denmark}

\author[0000-0001-9978-2601]{Aidan P. Cloonan}
\affiliation{Department of Astronomy, University of Massachusetts Amherst, Amherst MA 01003, USA}

\author[0000-0001-7160-3632]{Katherine E. Whitaker}
\affiliation{Department of Astronomy, University of Massachusetts Amherst, Amherst MA 01003, USA}
\affiliation{Cosmic Dawn Center (DAWN), Denmark}


\author[0000-0003-1641-6185]{Laia Barrufet}
\affiliation{Institute for Astronomy, University of Edinburgh, Royal Observatory, Edinburgh, EH9 3HJ}

\author[0000-0001-5063-8254]{Rachel Bezanson}
\affiliation{Department of Physics and Astronomy and PITT PACC, University of Pittsburgh, Pittsburgh, PA 15260, USA}

\author[0000-0003-3917-1678]{Rebecca A. A. Bowler}
\affiliation{Jodrell Bank Centre for Astrophysics, Department of Physics and Astronomy, School of Natural Sciences, The University of Manchester, Manchester, M13 9PL, UK}

\author[0000-0001-8460-1564]{Pratika Dayal}
\affiliation{Kapteyn Astronomical Institute, University of Groningen, 9700 AV Groningen, The Netherlands}

\author[0000-0002-8871-3026]{Marijn Franx}
\affiliation{Leiden Observatory, Leiden University, P.O.Box 9513, NL-2300 AA Leiden, The Netherlands}

\author[0000-0002-5612-3427]{Jenny~E.~Greene}
\affiliation{Department of Astrophysical Sciences, Princeton University, 4 Ivy Lane, Princeton, NJ 08544, USA}

\author[0000-0003-3760-461X]{Anne Hutter}
\affiliation{Niels Bohr Institute, University of Copenhagen, Jagtvej 128, Copenhagen, Denmark}
 \affiliation{Cosmic Dawn Center (DAWN), Copenhagen, Denmark}

\author[0000-0001-7673-2257]{Zhiyuan Ji}
\affiliation{Steward Observatory, University of Arizona, 933 North Cherry Avenue, Tucson, AZ 85721, USA}

\author[0000-0002-2057-5376]{Ivo Labb\'e}
\affiliation{Centre for Astrophysics and Supercomputing, Swinburne University of Technology, Melbourne, VIC 3122, Australia}

\author[0000-0003-0415-0121]{Sinclaire M. Manning}
\affiliation{Department of Astronomy, University of Massachusetts Amherst, Amherst MA 01003, USA}

\author[0000-0003-0695-4414]{Michael V. Maseda}
\affiliation{Department of Astronomy, University of Wisconsin-Madison, 475 N. Charter St., Madison, WI 53706, USA}

\author[0000-0003-1207-5344]{Mengyuan Xiao}
\affiliation{Department of Astronomy, University of Geneva, Chemin Pegasi 51, 1290 Versoix, Switzerland}

\begin{abstract}

We present the PANORAMIC survey, a pure parallel extragalactic imaging program with NIRCam observed during JWST Cycle 1. The survey obtained $\sim$530 sq arcmin of NIRCam imaging from 1-5$\mu$m, totaling $\sim$192 hours of science integration time. This represents the largest on-sky time investment of any Cycle 1 GO extragalactic NIRCam imaging program by nearly a factor of 2. The survey includes $\sim$432 sq arcmin of novel sky area not yet observed with JWST using at least $6$ NIRCam broad-band filters, increasing the existing area covered by similar Cycle 1 data by $\sim$60\%. 70 square arcmin was also covered by a 7th filter (F410M). A fraction of PANORAMIC data ($\sim$200 square arcmin) was obtained in or around extragalactic deep-fields, enhancing their legacy value.  Pure parallel observing naturally creates a wedding cake  survey with both wide and ultra-deep tiers, with 5$\sigma$ point source depths at F444W ranging from 27.8-29.4 (ABmag), and with minimized cosmic variance. The 6+ filter observing setup yields remarkably good photometric redshift performance, achieving similar median scatter and outlier fraction as CANDELS ($\sigma_{\rm NMAD}\sim0.07$; $\eta\sim0.2$), which enables a wealth of science across redshift without the need for followup or ancillary data. 
We overview the proposed survey, the data obtained as part of this program, and document the science-ready data products in the first data release. 
PANORAMIC has delivered wide-area and deep imaging with excellent photometric performance, demonstrating that pure parallel observations with JWST are a highly efficient observing mode that is key to acquiring a complete picture of galaxy evolution from rare bright galaxies to fainter, more abundant sources at all redshifts. 
\end{abstract}


\section{Introduction} \label{sec:intro}

In past decades, 
deep observations with the Hubble and Spitzer Space Telescopes traced the evolution and 
the assembly of galaxies up to $z\sim8$, and detected a handful of sources up to $z\sim11$ \citep[e.g.,][]{Song16,Davidzon17,Oesch18}. 
A hard boundary prevented pushing further into the early Universe due to Hubble's limited area and wavelength coverage, and Spitzer's  limited sensitivity to detect the red light of previously unexplored galaxy populations. Thus, many critical open questions remained prior to JWST launch, including:
When and how do the first galaxies form out of the primordial gas at z$>$9? When and how do the first galaxies shut off their star-formation and turn quiescent after $z\sim6$?  What fraction of the cosmic star-formation rate density is obscured and still missing at $z>3-7$?  All of these questions lie at the core of JWST's mission, necessitating deep imaging at 2-5\micron. 
While JWST/NIRCam's unprecedented sensitivity and resolution probes more than 100$\times$ fainter than Spitzer/IRAC at $3-5$\micron\   \citep{Rieke2023}, the above questions can only be answered with statistical samples of early massive galaxies. As massive galaxies become increasingly rare with redshift, the required areas to probe of the sky are significantly larger than the blank-field surveys that were planned by the GTO and ERS teams prior to Cycle 1 \citep[$\sim$ 430 square arcmin in total;][]{Windhorst2023,Finkelstein2023,Bagley2023,Eisenstein2023}. This motivated pursuit of wider area imaging among Cycle 1 GO programs.

The advent of the first JWST Cycle 1 datasets immediately revolutionized our view of galaxies in the early Universe even from small-area fields probed by the first datasets, and led to several surprising discoveries. 
The first NIRCam images revealed hitherto unknown galaxy populations, which are posing complex challenges to our understanding of galaxy formation. These include: (1) overly luminous galaxies at $z>10$ that appear more abundant than expected even from the most optimistic models \citep[e.g., ][]{Donnan22b,Harikane22,Atek23,Naidu22}. (2) extremely massive galaxy candidates whose early formation time could challenge our cosmological models \citep[e.g.][]{Labbe23a,BoylanKolchin22,Xiao23}. (3) Quiescent galaxies at $z\sim3.5-5$, which only further increases the tension between observations and theory \citep{Carnall22,Glazebrook2023, deGraaff2024, Weibel2024b}, requiring a stronger and earlier influence of AGN feedback \citep{Park2023, KurinchiVendhan2023}. (4) A significant population of dust-obscured, red galaxies at $z>3-9$, proving that our census of star formation with HST+Spitzer was incomplete \citep{Barrufet22,  Xiao23, Williams2023b}. (5) A puzzling high abundance of  obscured active galactic nuclei at $z\sim5-7$ (so-called little red dots, LRDs), whose black hole to stellar mass ratios put pressure on black hole seeding and growth models \citep{Labbe2023b,Kocevski2023,Greene24,Matthee23, Kokorev2023, Furtak23b}.

While these early JWST results could present daunting challenges to our models of galaxy formation, they are only based on small numbers of sources collected over a limited survey volume of current imaging datasets. Even 2 years into the JWST mission, the total area in extragalactic deep/blank-field imaging at NIR wavelengths (using 6 or more NIRCam filters), including GO programs, remains only $\lesssim$730 square arcmin\footnote{based on similar observing strategies to PANORAMIC. COSMOS-Web mapped 0.5 square degrees but only used 4 NIRCam filters \citep{Casey2023}.}  \citep[][]{Windhorst2023,Eisenstein2023,Finkelstein2023,Bagley2023,Dunlop2021}. To truly establish the number densities of these rare objects, significantly larger areas of the sky need to be surveyed \citep[e.g.][]{Moster2011,Trenti2008,Bhowmick2019,Ucci2021}. Further, the discovery of intrinsically bright sources that we need for spectroscopy must be selected from wider areas.

Unfortunately, JWST is not optimized for targeted wide-area surveys. While revolutionary for its sensitivity and spatial resolution, NIRCam's field of view, at $\sim$ 10 square arcmin in area, is of similar size to that of its predecessors on HST and Spitzer. 
The observing times required to cover wide areas (0.1-1 square degree scales) are dominated by overhead cost, and these observing strategies can also significantly reduce JWST's lifetime due to a build-up of momentum which results in high fuel consumption. While Roman and Euclid will ultimately provide wide-area mapping capabilities in space at wavelengths $<2\mu$m, JWST remains the only facility with $>$2-5$\mu$m coverage for the foreseeable future. The $>$2-5$\mu$m wavelength capability is the necessary tool enabling the discovery space of these poorly explored galaxy populations at $z>3$, whose red colors necessitate measurements at longer wavelengths.
To fully capitalize on JWST's unparalleled imaging and spectroscopic capabilities \citep{Gardner2023}, it is critical that we map large areas 
in order to pinpoint rare, yet critical evolutionary phases in the lives of early galaxies. Further, we must discover the most precious, intrinsically luminous candidate galaxies, since they are our best opportunities to study the detailed astrophysics of the first stars and galaxies with spectroscopy.

Fortunately, JWST is capable of pure parallel observations, building on decades of pure parallel work with Hubble that have established the wide-area tiers for extragalactic surveys.  
The successful HST pure parallel imaging programs (e.g. BORG and HIPPIES; \citealt{Trenti11, Trenti12, Yan11, Calvi16, Morishita18}) led the search for intrinsically bright galaxies during the reionization era, building our knowledge of the bright end of the UV luminosity function \citep{Bradley12, Schmidt14, RojasRuiz2020,  RobertsBorsani2022, Bagley2024}. JWST is now demonstrating the success of these programs by confirming the high-redshift nature of these intrinsically bright discoveries at $z\sim7-9$ \citep{RobertsBorsani2024, RojasRuiz2024}. However, the multiplex capabilities of JWST instruments provides exciting advances beyond what was possible with HST. While HST pure parallel observations typically only obtained a 3-filter baseline design for Lyman-break color selection, science with this data often required substantial Hubble and Spitzer imaging follow-up to obtain more extensive multi-wavelength data. Since NIRCam simultaneously images in the short- and long-wavelength channel made possible by its dichroic \citep{Rieke2023}, it is particularly efficient at building up multi-wavelength photometry in a single visit. JWST's ability to accumulate deep imaging in parallel is also unprecedented: the equivalent area of a compilation of parallel imaging with HST over 10+ Cycles is achievable by JWST in one Cycle \citep[e.g. 0.4 square degrees, SuperBORG;][see Section \ref{sec:realcycle1}]{Morishita2021}.

\begin{figure*}
    \centering
\includegraphics[width=0.85\textwidth]{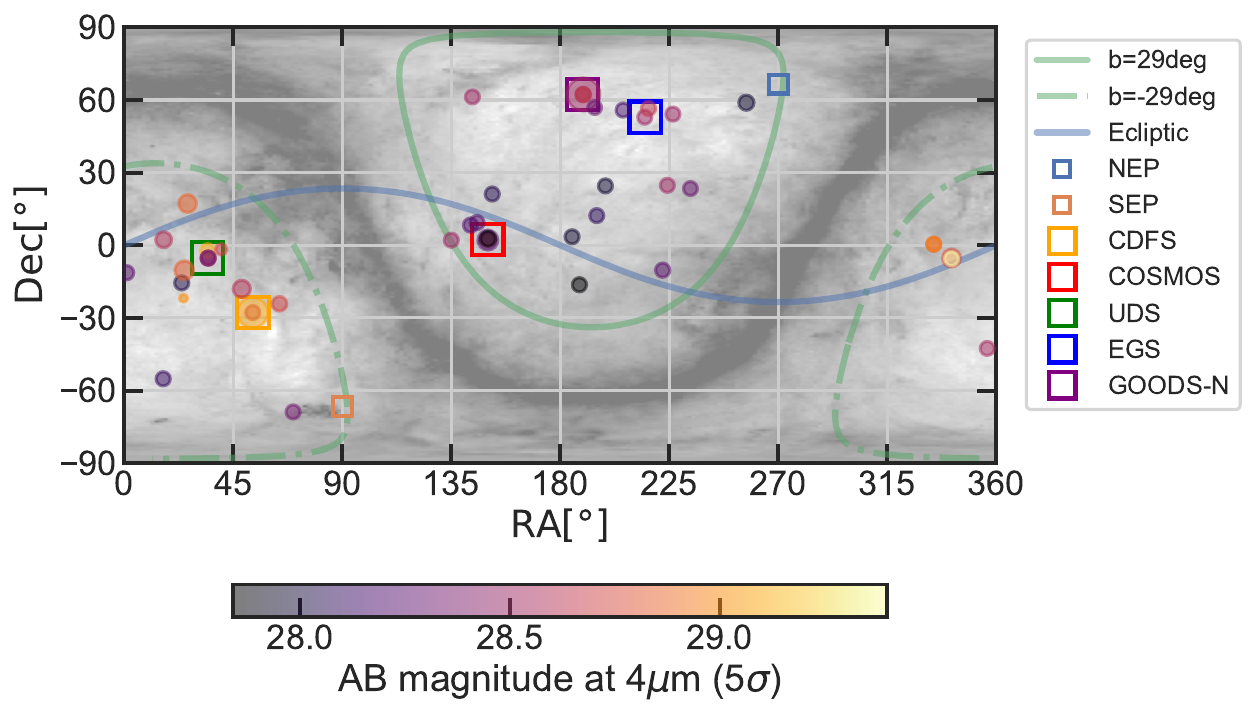}
\caption{Full sky map of our observed pointings. Circles indicate the location of data obtained by PANORAMIC, color coded according to the 5$\sigma$ detection limits at 4$\mu$m, with circle size scaled by on-sky area (not to scale). Existing extragalactic fields are marked with squares, and CANDELS fields (where we obtained data) are also additionally exaggerated in size for visibility. Background grayscale indicates the 4$\mu$m background, with regions of low background (galactic latitude $|b|>$29) used to select parallel visits indicated by green solid and dot-dashed lines. }
    \label{fig:map}
\end{figure*}

While pure parallel observations allow efficient mapping of wide areas on the sky with negligible cost to observatory resources, observing non-contiguous pointings is even more beneficial for rare galaxy science. Mapping uncorrelated areas of the sky enables the detection of overdensities at the highest redshifts most efficiently, minimizing cosmic variance  \citep[e.g. simulations for JWST pure parallels;][]{Trapp22a}.
Especially when searching for the rare, luminous galaxies that are expected only in overdense regions, cosmic variance can be a dominant contribution to the uncertainty budget \citep{Trenti2008, Trapp2020,  Jespersen2024}. While most lines of sight will not contain any galaxy, the densest regions may contain many sources \citep{Steinhardt2021}, motivating the collection of as many independent sightlines as possible. 
Thus, by design, pure parallel JWST surveys represent the most efficient way to pinpoint the most overdense regions of the Universe where galaxy formation started first, and provide the most efficient data for follow-up studies.

This paper introduces the PANORAMIC Survey (Parallel wide-Area Nircam Observations to Reveal And Measure the Invisible Cosmos), the first extragalactic pure parallel imaging program executed using JWST in Cycle 1 (JWST-GO-02514; PIs: C. Williams and P. Oesch).
Thanks to a flexible and tiered, $\geq$6-filter survey strategy, this program was capable of efficiently using many heterogenous parallel opportunities to map wide areas. 
Thanks to the dichroic enabling simultaneous imaging in 2 filters, the NIRCam parallel images obtained as part of PANORAMIC are fully self-contained, providing reliable samples and accurate photometric redshifts (see section \ref{sec:eazy}) resulting in an enormous legacy dataset for the community. Pure parallels are a powerful (and needed) wide-tier dataset that complement targeted deep field programs: the data obtained as part of this program naturally follow the ``wedding cake" survey strategy, including both wide-area and ultra-deep tiers, that have become a gold standard strategy for extragalactic surveys \citep[e.g. CANDELS;][]{Grogin11}. In Section \ref{sec:design} we overview the proposed observations and survey design, and the actual Cycle 1 observations. In Section \ref{sec:data}, we present our data processing methods including image reduction, photometric measurements, redshift estimations that make up the initial data release. In Section \ref{sec:science} we overview the science goals that motivated the PANORAMIC survey and highlight the exciting science potential of pure parallel imaging with JWST. We assume a $\Lambda$CDM cosmology with  H$_0$=70 km s$^{-1}$ Mpc$^{-1}$, $\Omega_M$ = 0.3, $\Omega_\Lambda$ = 0.7, 
and magnitudes are specified in the AB system \citep{Oke83}.

\section{Survey Design \& Cycle 1 Observations} \label{sec:design}

The pure parallel observing mode by nature is dependent on ``prime-instrument" observational designs (i.e. the telescope configuration while collecting data on primary targets of other programs approved in the same Cycle). The areas and integration times that are achievable in parallel cannot be known a priori. Thus, pure parallel observations are considered shared risk (not guaranteed) and the realized Cycle 1 dataset naturally differs from that proposed. In the sections below, we first introduce the as-proposed survey design (Section \ref{sec:designprop}, which we used as a baseline for optimizing the designs of the real observations). Then, in the following Section \ref{sec:realcycle1} we describe the realized dataset.

\begin{figure}
    \centering
   \includegraphics[scale=.5]{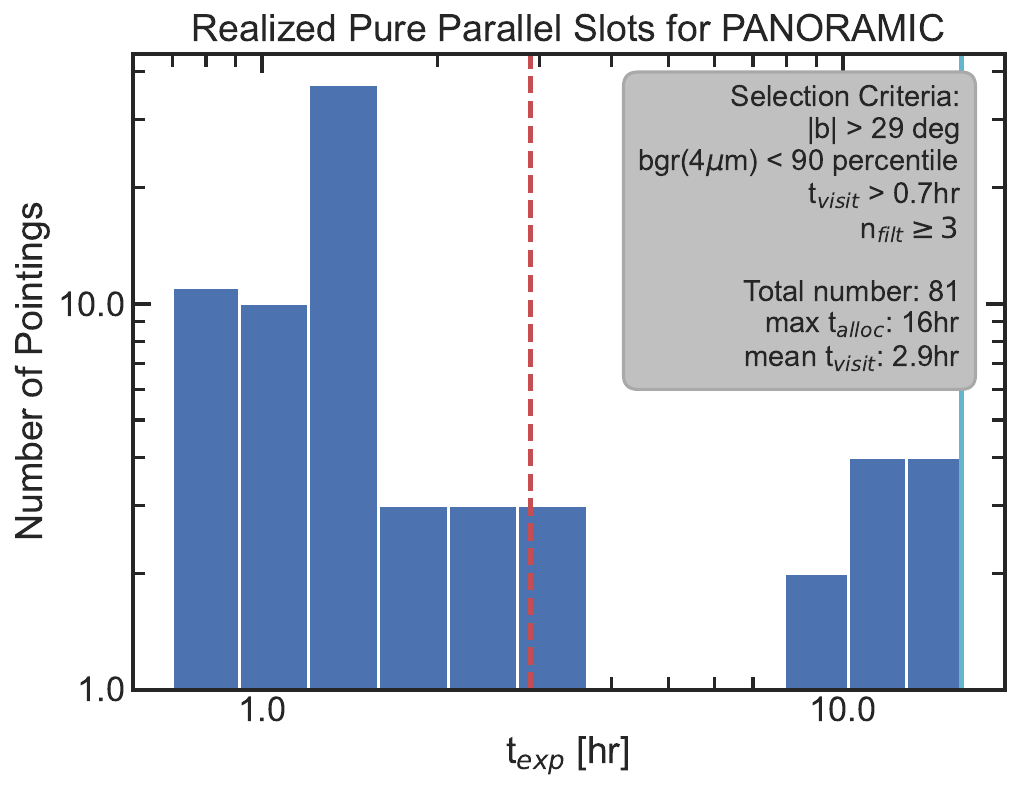}
    \caption{Distribution of the total exposure times per individual pointing for NIRCam parallels that were designed in as part of this program, to illustrate the capability for NIRCam parallels during Cycle 1 (not all designs were observed; see Appendix \ref{sec:mishaps}). The PANORAMIC parallel visits were selected to have long total integration times ($>$42 minutes), high galactic latitude $| b | >29$,  low 4$\mu$m background, and have at least 2 mechanism moves (3 filter-sets). While we requested visits with at least 1 hour total integration for our minimum survey design, the real distribution of integration times per visit are typically longer (on average 2.9 hours; red dashed line; median 1.5 hours). To avoid data loss due to data rate limitations, we found that the parallel-eligible visits with long total exposure times had to be truncated to $\sim16$ hours if NIRCam was the parallel instrument (cyan line). } 
    \label{fig:nexpdist}
\end{figure}

\subsection{Proposed survey design}\label{sec:designprop}

The PANORAMIC survey design (i.e., the NIRCam filters, depths, and number of NIRCam pointings) is driven by the requirement to probe a statistical sample of relatively bright (F200W $\lesssim $27 ABmag) galaxies out
to the highest redshift, and to create the wide-area tier of JWST extragalactic imaging for the widest array of science.
The baseline strategy mandates at least 1 hour on-source time, split over the 3 short-wavelength filters F115W, F150W and F200W, and the three long-wavelength filters F277W, F356W, and F444W. For 20 minute exposures, the expectation from the exposure time calculator (ETC) was
5$\sigma$ depths of 27.6-28.3 ABmag in apertures of 0\farcs16 radius, under medium background conditions.  Our proposed area was 150 pointings, approximately 0.4 square degrees ($\sim$1500 sq arcmin; see the next section for the actual achieved survey area).  This remarkably wide area is comparable to the widest Cycle 1 program COSMOS-Web \citep{Casey2023}, but would have effectively tripled the deep extragalactic survey area from ERS+GTO surveys while allowing $\geq6$ NIRCam filters, at essentially zero overhead cost.

Given that the pure-parallel slots are assigned to a primary
program, the achieved data includes many longer visits than the minimal request of 1 hour. As was the case for HST parallels \citep[e.g.,][]{Trenti11, Yan11, WISPS, Calvi16}, we find that the real JWST observations in any given Cycle follow a power-law distribution of exposure time. 
Therefore, any pure-parallel survey has a `built-in' multi-tiered design with a large area covered by relatively shallow pointings and a small number of very deep pointings. While the prime-instrument always drives the division of exposure time between filters, where possible, we planned to split the exposure times up to obtain the deepest dataset in F115W, which significantly reduces the outlier fractions in photometric redshifts. In this case, the least
efficient filter, F444W, shares this deeper slot. When more than 3 mechanism moves (i.e., filter wheel changes) are allowed by the primary program, we additionally obtain a fourth filter-set including the medium band filter F410M (paired with additional imaging in F115W to increase depth), motivated to improve the photometric redshifts and physical constraints for fainter red galaxies at $3 < z < 7$ \citep[e.g.][]{Kauffmann2020,RobertsBorsani2021}.

\subsection{Cycle 1 observations}\label{sec:realcycle1}

The data described here was obtained as part of Cycle 1 program ID 2514 (PIs C. Williams \& P. Oesch). Due to unanticipated pressure on the deep space network (DSN) during JWST commissioning from January-June 2022 \citep[][]{Rigby2023}, the start of science pure parallels in Cycle 1 was postponed by three months in order to first gain experience during science operations with the DSN connectivity before implementing science pure parallels (which increase the overall data volume obtained by the spacecraft).   
Due to this delay, in addition to various software technical difficulties that were unforeseen, JWST has only been executing science pure parallel observations since December 2022\footnote{https://parallels.stsci.edu/jwst/1/status-report.html}. Thus, science pure parallel data was collected for only approximately half of Cycle 1.

The proposed survey design required at least 6 filters, with integration times that total $>$1 hour across all filters, which correspond to $\gtrsim$ 28 ABmag (5$\sigma$ detection limit) in F444W. 
Due to the late start of science pure parallel observations, we relaxed the minimum requirement of 1 hour integration per pointing to 42 minutes to increase the area that could obtain data for the remainder of Cycle 1. 

The observational design of PANORAMIC can be achieved through a range of possible prime instrument observational setups. Thus, we selected from available pure parallel slots for this survey according to the following criteria: First, we required that parallel slots allow 2 or more mechanism moves. This allows NIRCam to change filters twice, resulting in imaging for 3 filter sets, i.e., 6 filters total when accounting for the fact that NIRCam can image in the short- and long-wavelength channel at the same time. 
Furthermore, we required that the prime instrument's integration time for individual exposures allow for NIRCam readout patterns with at least 4 groups for ramp fitting, to ensure robust cosmic ray rejection. This means that we only selected parallel opportunities where the minimum individual exposure time was 182 seconds (4 groups with SHALLOW2 readout). All of our pointings were designed with at least 2 exposures per filter (corresponding to 2 dither locations). We additionally required that observations be taken at low background conditions at 4$\mu$m (10th percentile) and that pointings lie well outside the plane of the Milky Way at galactic latitude +/- 29 degrees or higher/lower. For assymetric division of integration time per filter, we paired the longest integrations with the least sensitive filters. Finally, we visually inspected the prime-instrument on-sky footprint for each dither-set (i.e., the exposures that would be taken with the same parallel filter) within a visit, and between visits, of the same prime program. This step was necessary to determine whether each of the 3 filter-sets mapped in parallel by NIRCam would result in a majority overlapping area between the 6 filters. In some cases this allowed us to group exposures from different visits together that map similar area of the sky, in order to increase the integration time and number of filters.  The on-sky locations of our obtained data is presented in Figure \ref{fig:map}.

As shown in Figure \ref{fig:nexpdist}, the real distribution of integration times among pure-parallel-eligible observations for NIRCam in Cycle 1 that met the above criteria include a range of total exposure times, from 42 minutes to a small number of very deep pointings with $\sim$30 hours. However, in practice, we found that it is not possible to safely design NIRCam parallel observations that take advantage of the full integration among those very long parallel slots, without risk of exceeding the data rate limitations of the solid state recorder onboard JWST. The actual data volume depends on the data rate of other programs scheduled between DSN connections and thus cannot be known in advance at the time of the APT designs. Thus, we opted for a conservative approach of not exceeding the minimum data volume recommendations\footnote{\url{https://jwst-docs.stsci.edu/jwst-general-support/jwst-data-volume-and-data-excess}} and found that this typically limited the pure-parallel NIRCam data collection to $\sim$16 hours per pointing (corresponding to $<4$ hours per filter, because the prime instruments typically allowed 4 mechanism moves during longer visits).

\begin{figure}[!t]
    \centering
    \includegraphics[scale=.6]{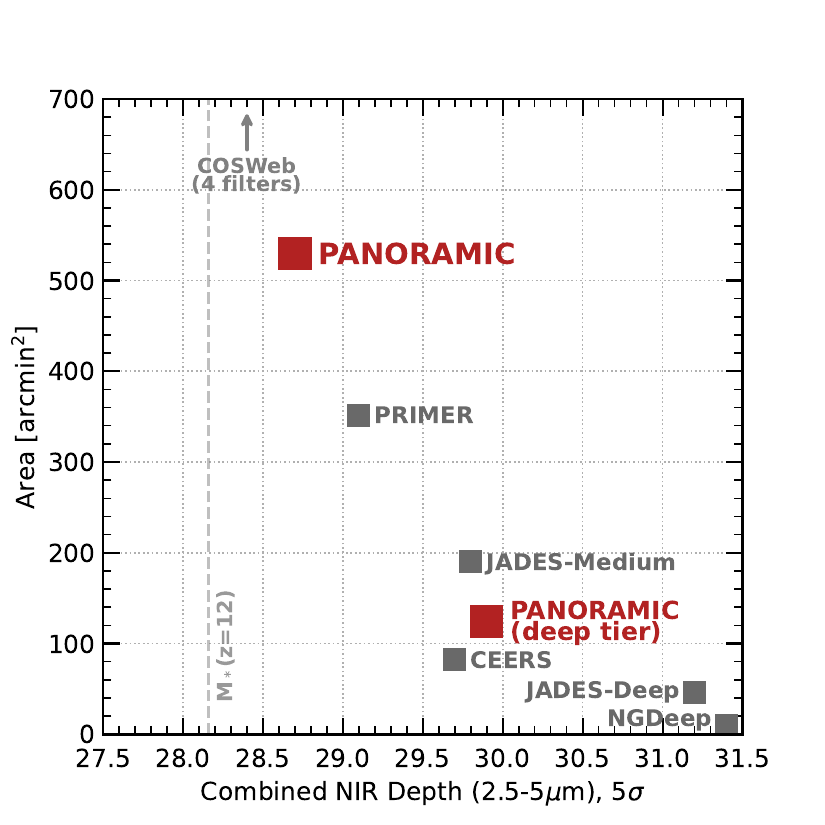}
    \caption{Area and depth of the PANORAMIC survey relative to other cycle one programs. The near-infrared depth was estimated by a weighted combination of all filters in the long-wavelength channel of NIRCam, i.e., covering $\sim2.5-5\,\mu$m. The top 20\% deepest pointings of PANORAMIC combined (deep tier) are of comparable depth and area as the ERS CEERS survey. The full area is shown at the depth of the 20\% shallowest pointings (28.7 mag). For reference, the characteristic magnitude of the UV luminosity function at $z\sim12$ is shown as dashed vertical line. }
    \label{fig:areadepth}
\end{figure}

Scheduling of parallel data for PANORAMIC began in January 2023 and the first data arrived in February 2023. In March 2023, a series of problems with updates to the scheduling software resulted in a series of pointings obtained between March-April 2023 only having 4 filters (see Appendix Section \ref{sec:mishaps}, where we overview these various issues). By the end of Cycle 1, the program obtained 55 pointings ($\sim$ 530 square arcminutes) with at least 2 filters of imaging. Our program successfully obtained 45 pointings (corresponding to $\sim$432 sq arcmin, or 0.12 sq degrees) using our $\geq$6 filter setup as initially proposed. This effectively increases the available extragalactic imaging with $\geq6$ filters from Cycle 1 by 60\% (compared to $\sim730$ square arcminutes from ERS, GO and GTO combined; see Figure \ref{fig:areadepth}). 
Approximately 200 square arcminutes of PANORAMIC data falls near or around extragalactic legacy fields, and 40 associations (45 pointings) probe novel area with JWST (in our NIRCam filters) for the first time. An overview of the imaging that PANORAMIC obtained is listed in Table \ref{tab:nircam}, and a summary of the pointings that contribute to existing imaging in extragalactic legacy fields are presented in Table \ref{tab:exgal}.

We finish this section by noting that while the actual scheduling efficiency for pure-parallel NIRCam observations was essentially unknown prior to approval of Cycle 1 GO programs, our experience offers a first opportunity to explore the typical parallel designs that are feasible within one cycle. 
Although science pure-parallels  started 6 months late in Cycle 1 (about half-way through, excluding scheduling rollover) our program was successful at designing about half of the requested parallel pointings (81 out of 150), 
all of which met the proposed criteria ($\gtrsim42$ minutes total integration with $\geq$6 filters; see public APT file). While not all of these observations were observed for various reasons (see Appendix Section \ref{sec:mishaps}), the fact that we could design half of the pointings to fit within half of a cycle nonetheless confirms that for a typical cycle, it is feasible for NIRCam parallels to collect $\sim$0.4 square degrees while using 2 mechanism moves (3 filter-sets). 
Since the various technical problems that prevented data collection described in Appendix \ref{sec:mishaps} are now resolved, it is reasonable to expect that future cycles can obtain similar areas to the originally proposed $\sim$0.4 square degrees of imaging in parallel using 3 filter-sets.  
We find that only $\sim$10\% of these parallels (7 pointings) can accommodate a 3rd mechanism move (4 filter-sets), which represents a relatively limited area. This highlights the importance of maintaining flexibility with a tiered parallel design to enable successful collection of observations over wide-areas.

\begin{deluxetable*}{llllllll}[!htbp]
\caption{44 associations obtained by PANORAMIC for which we release image mosaics (excludes pointings where footprints entirely overlap deeper data in EGS, COSMOS-PRIMER, and UDS-PRIMER; these are recorded in Table \ref{tab:exgal} and in Figure \ref{fig:legacypointings}).  }\label{tab:nircam}
\tablehead{\colhead{Association}  & \colhead{Field$^{a}$}&  \colhead{N$_{exp}$} &  \colhead{N$_{\rm filters}$}& \colhead{RA} & \colhead{Dec}  & \colhead{Obs date$^{b}$} & \colhead{E(B-V)} }
\startdata
j000352m1120 &         & 120 & 6 & 0.960430  & -11.326817  & 2023-07-14   &  0.029 \\
j010408m5508 & 	       & 120 & 6 & 16.039404 & -55.141650  & 2023-08-05   &  0.016 \\
j010500p0217 & 	       & 200 & 7 & 16.245068 & 2.282488    & 2023-08-18   &  0.023 \\
j013444m1532 & 	       & 120 & 6 & 23.687083 & -15.531187  & 2023-08-13   &  0.019 \\
j013748m2152 & MRG0138 & 160 & 7 & 24.453443 & -21.864537  & 2023-12-27   &  0.013 \\
j013900m1019 & 	       & 180 & 6 & 24.743085 & -10.320011  & 2023-08-23   &  0.019 \\
j014428p1715 & 	       & 180 & 6 & 26.108409 & 17.240552   & 2023-08-29   &  0.050 \\
j021716m0520 & UDS     &  70 & 6 & 34.323502 & -5.331135   & 2023-08-20   &  0.023 \\ 
j021728m0214 & XMM/UDS & 240 & 6 & 34.362961 & -2.228235   & 2023-08-06   &  0.020 \\
j021800m0521 & UDS     &  70 & 6 & 34.496029 & 	-5.354210   & 2023-08-19   &  0.019 \\ 
j021824m0517 & UDS     & 70  & 6 & 34.598771 & -5.278195   & 2023-07-23   &  0.018 \\
j024000m0142 &Abell370 & 90  & 6 & 40.007217 & -1.691671   & 2023-09-05   &  0.026 \\
j031404m1759 & 	       & 180 & 6 & 48.511560 & -17.973015  & 2023-09-17   &  0.027  \\
j033212m2745 &GDS      &  70 & 6 & 53.009876 & -27.765500  & 2024-01-27   &  0.008 \\
j033224m2756 & GDS &  460  & 7 & 53.070160  & -27.946099       & 	2023-10-14    	 &  0.006 \\
j041616m2409 & MACS0416 & 210 & 6 & 64.090215 & -24.179155 & 2023-08-27   &  0.035 \\
j043844m6849 & LMC     &60   & 6 & 69.682710 & -68.821463  & 2023-07-25   &  0.079 \\
j090000p0207 & 	       & 80  & 4 &135.003870 &  2.114068   & 2023-04-03   &  0.037 \\
j093144p0819 & 	       &120  & 6 &142.941080 &  8.325030   & 2023-04-20   &  0.039 \\
j093452p6116 & 	       &80   & 4 &143.708385 & 61.272367   & 2023-04-06   &  0.029 \\
j094232p0923 & 	       &120  & 6 &145.629850 &  9.378996   & 2023-04-19   &  0.024 \\
j100024p0208 &COSMOS-WEB  &140  & 6 &150.104485 &  2.131158   & 2024-01-06   &  0.016 \\
j100040p0242 & COSMOS &120   & 6 &150.164400 &  2.693489   & 2023-12-03   &  0.016 \\
j100736p2109 & 	      & 60   & 6 &151.906160 & 21.155276   & 2023-05-07   &  0.025 \\
j121932p0330 & 	      &120   & 6 &184.880170 &  3.497880   & 2023-06-12   &  0.016 \\
j123140m1618 & 	      &120   & 6 &187.911310 & -16.305448  & 2023-07-02   &  0.037 \\
j123732p6216 &GDN & 180  & 6 & 189.380245 &  62.259910  & 2024-05-19   &  0.010 \\
j123744p6211 & GDN    &70    & 6 &189.438515 & 62.180307   & 2024-01-30   &  0.010 \\
j123816p6214 & GDN    & 90    & 6 &189.560465 &  62.232738  & 2023-03-24   &  0.010 \\
j125652p5652 & 	      &240   & 6 &194.211875 & 56.865695   & 2023-04-01   &  0.009 \\
j130016p1215 & 	      & 60   & 6 &195.064890 & 12.249040   & 2023-06-25   &  0.024 \\
j131432p2432 & 	      &120   & 6 & 198.635240 & 24.529744  & 2023-06-11   &  0.012 \\
j134348p5549 & 	      &120   & 6 & 205.941740 & 55.819387  & 2023-06-10   &  0.008 \\
j142536p5630 & 	      &120   & 6 & 216.404605 & 56.507497  & 2023-06-24   &  0.014 \\
j144904m1017 &        &120   & 6 & 222.271220 &-10.283259  & 2023-07-02   &  0.099 \\
j145652p2444 & 	      &40    & 4 & 224.223640 & 24.740027  & 2023-03-24   &  0.036 \\
j150604p5409 & 	      &160   & 4 & 226.514625 & 54.154775  & 2023-04-03   &  0.014 \\
j153500p2325 & 	      &240   & 6 & 233.742220 & 23.420023  & 2023-03-06   &  0.045 \\
j170720p5853 & 	      &120   & 6 & 256.841610 & 58.879343  & 2023-02-18   &  0.025 \\
j221648p0025 & SSA22  &120   & 7 & 334.169740 &  0.416990  & 2023-11-29   &  0.053 \\
j221700p0025 & SSA22  &120   & 7 & 334.246045 &  0.415294  & 2023-11-21   &  0.049 \\
j224604m0518 & 	      &180   & 6 & 341.515745 & -5.304581  & 2023-07-17   &  0.030 \\
j224616m0531 & 	      &430   & 7 & 341.565315 & -5.519552  & 2023-07-16   &  0.028 \\
j234512m4235 & 	      &80    & 7 & 356.304565 &-42.590595  & 2023-07-28   &  0.012 \\
\enddata
\tablecomments{ (a) We identify the name of the region if the association is within or adjacent to known field targets. (b) Observation date indicated is for the last image obtained.}
\end{deluxetable*}

\section{Data processing} \label{sec:data}

\subsection{Image reduction}\label{sec:reduction}

The basic data reduction, as well as the astrometric alignment, co-adding and mosaicking is performed using the grism redshift and line analysis
software for space-based spectroscopy (\texttt{grizli}) pipeline (G. Brammer in prep., version 1.9.13.dev26).

Using the stage 1 reduced NIRCam images from the Space Telescope Science Institute (STScI) stage 1 pipeline, \texttt{grizli} performs the masking for various artifacts, such as ``snowballs'', 1/f noise, and scattered light \citep[e.g. claws, wisps;][]{Rigby2023}. We use the updated \texttt{snowblind} code{\footnote{\url{https://github.com/mpi-astronomy/snowblind}}} for masking NIRCam snowballs and an updated NIRCam bad pixel mask that is included in the \texttt{grizli} code repository.

Astrometric alignment is performed in two steps. In fields with no prior JWST or HST data, the F444W images are first aligned to the \textit{Gaia} reference frame, either using the Gaia DR3 sources themselves \citep{GaiaDR3}, or bootstrapped from the deeper wide-field, ground-based Dark Energy Spectroscopic Instrument (DESI) Legacy Imaging Surveys\footnote{\url{https://www.legacysurvey.org/}} \citep[][and references therein]{Dey2019} catalog that itself is aligned to \textit{Gaia} but has a significantly higher source density. All of the remaining filters in a particular PANORAMIC field are then aligned to a source catalog extracted from the F444W image of that field.

All \texttt{grizli} pipeline settings correspond to the version 7.2 mosaic release hosted and described in additional detail on the DAWN JWST Archive (DJA; \url{https://dawn-cph.github.io/dja/imaging/v7/}; unless otherwise noted). In addition, we add the following aesthetic corrections to all data. We model and subtract out large diffraction spikes from bright stars at $\pm$30 and $\pm$90 degrees relative to the detector pixel axes. In the case of two pointings, the use of only 4 groups resulted in visibly worse cosmic ray removal. 
For associations \texttt{j153500p2325} and \texttt{j125652p5652} where the short exposures have too few samples to effectively identify cosmic rays in the individual exposure ramp fits, we use more aggressive AstroDrizzle parameters \texttt{driz\_cr\_snr} 1.0 0.5 and \texttt{driz\_cr\_scale} 1.0 0.6 to better identify cosmic rays from the ensemble of exposures.

\subsection{Image associations}

While the typical PANORAMIC pointing consists of just a single NIRCam footprint with some dithers, there are a few fields in which our program obtained several pointings from different pure parallel slots. We thus group different exposures into `associations' on the sky by searching around each within a buffer of 25 arcmin. 
This leads to several associations that include more than one pointing. 
In a second step, we manually regroup a few of these collections of pointings into different associations in order to, e.g., split out pointings with only 4 filter exposures, or to separate out exposures that fully overlap with existing legacy field data that are publicly released through the DJA archive.

\begin{figure*}[t]
\includegraphics[width=1\textwidth]{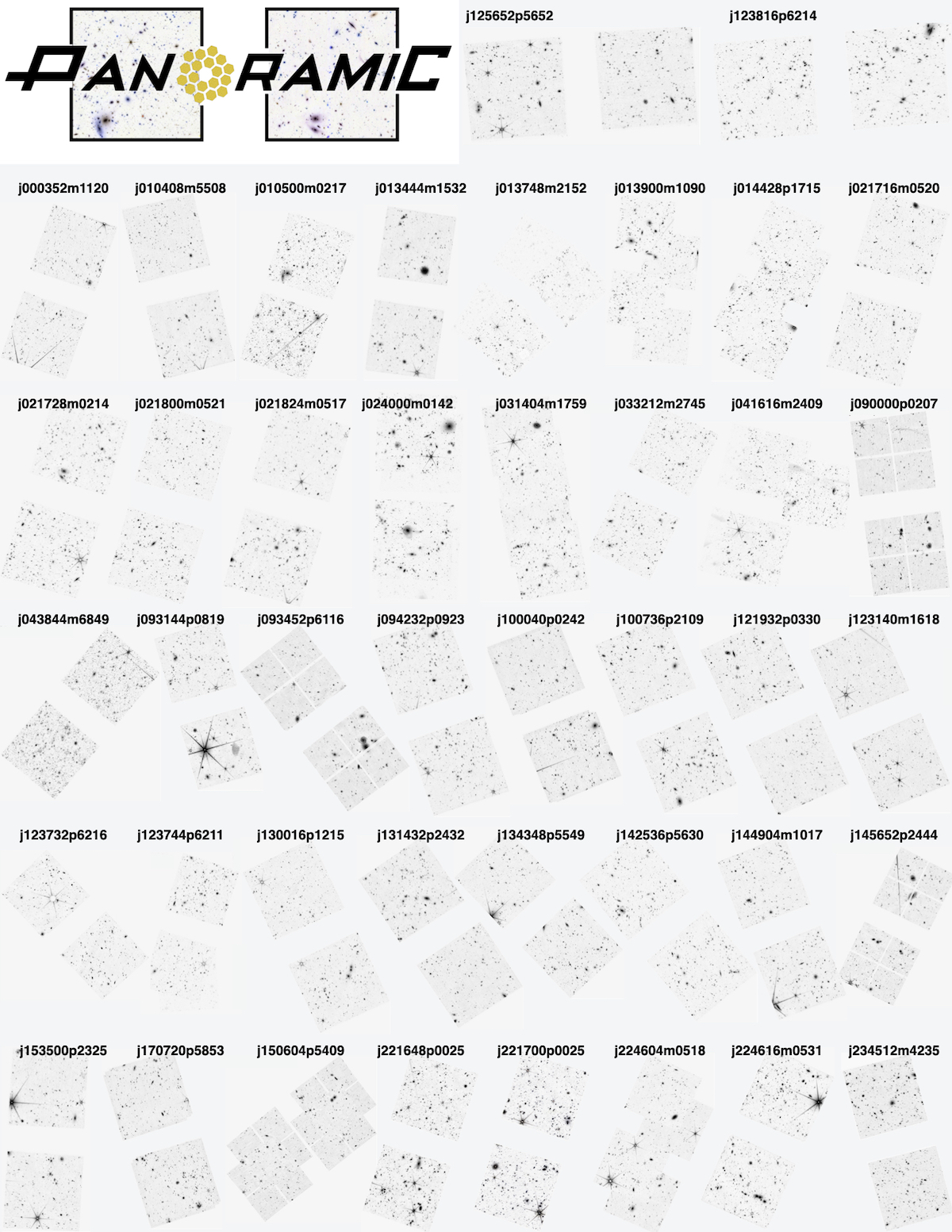}
\caption{The subset of  parallel imaging footprints from Table \ref{tab:nircam} that cover novel sky area with JWST (not all on the same scale). Large mosaics including substantial ancillary data are excluded for clarity (j033212m2745 in GOODS-S and j100024p0208 in COSMOS-Web; see Figure \ref{fig:legacypointings}). }\label{fig:novelpointings}
\end{figure*}

The associations are then named by their coordinates using ID \texttt{j[ra]p/m[dec]} where \texttt{[ra]} and \texttt{[dec]} are the mean coordinate of all exposures within the association in hours, minutes, seconds for ra and degrees, minutes for dec, respectively. 
For each association, we then define an output WCS frame to which all of the exposures are combined and drizzled using astrodrizzle \citep{Gonzaga2012}, with pixel scales of 20mas for the SW detectors and 40mas for the LW detectors. 
Information on the different associations and their images are listed in Tables \ref{tab:nircam} and \ref{tab:exgal}, where we also list the names of the known fields that are either adjacent to or overlapping with PANORAMIC. Images of fields with newly covered area are shown in Fig. \ref{fig:novelpointings}.

\begin{deluxetable*}{lllllllll}[htbp]
\caption{All pointings inside or proximate to extragalactic legacy fields}\label{tab:exgal}
\tablehead{\colhead{Field name}  & \colhead{N$_{pointings}$} &  \colhead{N$_{filters}$ }& \colhead{Association ID}  & & \colhead{Area [sq arcmin]}  }
\startdata
GOODS-S & 7 & 6-7 & j033212m2745  & & $\sim$60 \\
& & & j033224m2756 & & & & \\
& & & j033216m2755$^{a}$ &  & & &  \\
GOODS-N & 11 & 2-6 & j123744p6211 & & $\sim$80 \\
& & &  j123816p6214 &  & & & \\
& & &  j123732p6216 &  & & &  \\
& & & j123800p6217$^{a}$ &  & & &  \\
& & & j123816p6217$^{a}$ &  & & &  \\
& & & j123720p6215$^{a}$ &  & & &  \\
COSMOS & 1 & 6 & j100040p0242 & & $\sim$9 \\
COSMOS-Web & 2 & 6 & j100024p0208 & & $\sim$18 \\
COSMOS-PRIMER & 2 & 6 & j100028p0215$^{a}$ & & $\sim$18  \\
EGS & 1 & 6 & j141932p5254$^{a}$ & & $\sim$9  \\
UDS-PRIMER & 2 & 6 & j021748m0516$^{a}$  & & $\sim$18 \\
UDS & 3 & 6 & j021716m0520  & & $\sim$28 \\
& & & j021800m0521  &  & & &  \\
& & & j021824m0517  &  & & &  \\
XMM/UDS & 1 & 6 & j021728m0214 & & $\sim$9 \\
 &  & & & & \\
\enddata
\tablecomments{ (a)  associations are excluded from this data release (either because it sits entirely inside deeper data from other surveys, was corrupted by artifacts or NIRSpec shorts, or lacks full 6-filter coverage). Usable data will be folded into DJA mosaic releases coadded with other data.  }
\end{deluxetable*}

\subsection{Ancillary data}

\begin{figure*}[t]
\includegraphics[width=\linewidth]{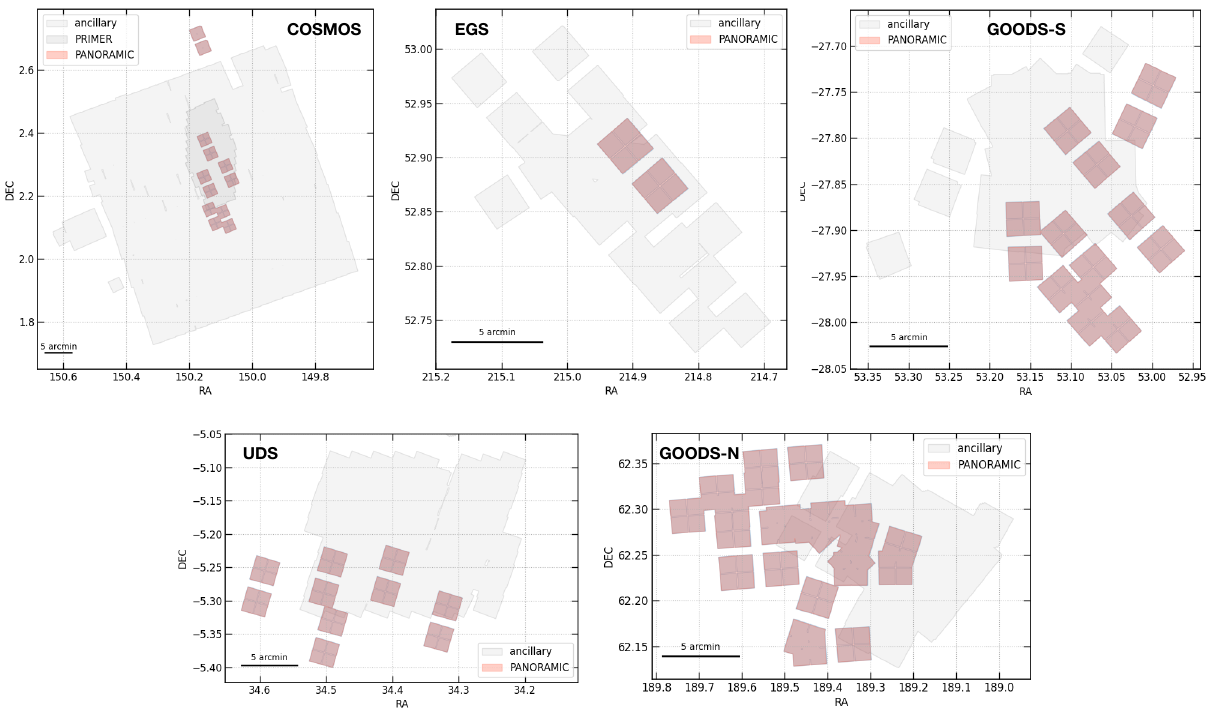}
\caption{Parallel imaging footprints in and around extragalactic deep fields. Gray regions show JWST footprints from prime-instrument programs 
that were publicly available as of July 1, 2024 \citep[including COSMOS-Web, PRIMER, JADES, and CEERS;][]{Casey2023, Eisenstein2023, Dunlop2021, Bagley2023}. Dark red indicates imaging obtained with PANORAMIC (PID 2514). 
PANORAMIC footprints show imaging with F444W (red) and F150W (blue outline) while gray ancillary footprints indicate coverage with any NIRCam filter. }
\label{fig:legacypointings}
\end{figure*}

While  PANORAMIC  was developed specifically to be useful as a stand-alone survey not relying on ancillary data, we still include all available HST and JWST imaging that overlaps with our association footprints in our analyses. This is done through the \texttt{aws} module of the \texttt{grizli} tool, which allows us to include all available reduced and aligned exposures in the archive. This is especially important in pointings that lie around legacy fields which often have  multi-wavelength coverage that extends beyond the legacy footprint due to parallel observations. The PANORAMIC pointings that lie near or overlapping with legacy fields are shown in Fig. \ref{fig:legacypointings}.

\subsection{A special pointing}

The above description of the PANORAMIC survey data applies to all pointings except for one: association j123732p6216 in the GOODS-N field.
This pointing was not initially available as a pure parallel eligible slot because it had a coordinated parallel, but became parallel-eligible after the prime-instrument observation required a repeat and the coordinated parallel data was successfully obtained. The orientation angle was fixed and known at the time of our parallel design in APT, and our footprint was to fall within extremely deep JADES imaging that already obtained all of our broad and medium band filters \citep{Eisenstein2023}. In this case we adapted our filter set to obtain 6 medium band filters (F162M, F182M, F210M, F300M, F430M, F460M) rather than incrementally adding to existing filters. The filter choice followed the motivation for existing medium band surveys such as the JWST Extragalactic Medium-band Survey (JEMS) and JADES Origins Field (JOF) survey \citep{Williams2023a, Eisenstein2023b}.

\subsection{Photometric catalogs}\label{sec:cats}

From the drizzled mosaics (Section \ref{sec:reduction}) we generate photometric catalogs, largely following the methods outlined in \citet{Weibel24}. In brief, we run \texttt{SourceExtractor} \citep{Bertin96} in dual mode, using an inverse-variance weighted stack of all the LW wide filters, F277+F356W+F444W as the detection image. We then measure fluxes in all the available individual filters through circular apertures of various radii, adopting a radius of 0.16\arcsec\ for our fiducial measurements. The images are PSF-matched to F444W prior to those measurements. As opposed to \citet{Weibel24}, we do not construct empirical PSFs from the stars in the images. Due to the small area of each pointing, the number of suitable stars is small in some pointings, making the PSF extraction difficult. Therefore, we derive PSFs with \texttt{webbpsf} \citep{Perrin2014} setting the \texttt{jitter\_sigma} parameter to 0.02 and rotating them to match the position angle of each pointing respectively.

\begin{figure*}
\centering
\includegraphics[]{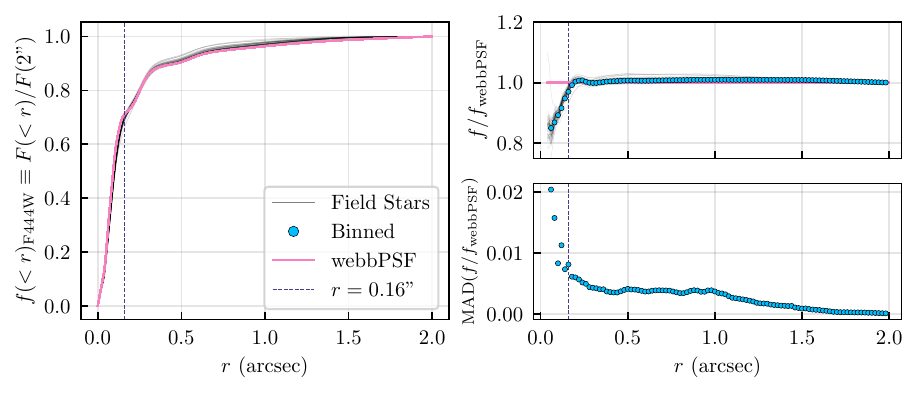}
\includegraphics[]{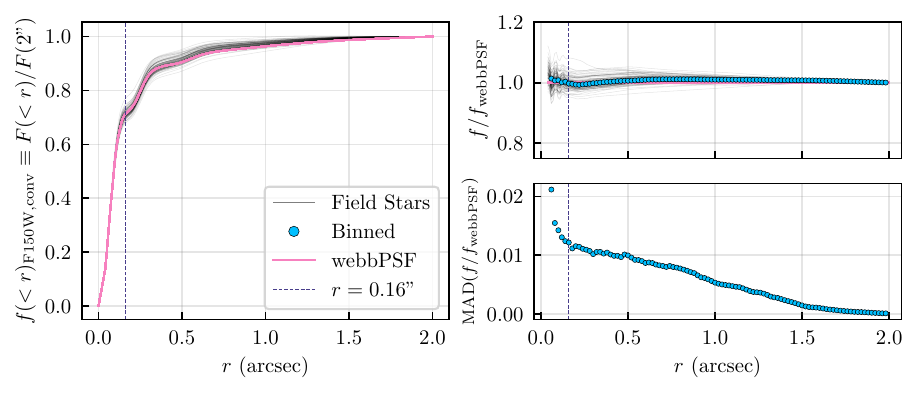}
\caption{An analysis of the PSF curve of growth (CoG) with radius $r$ for F444W (top) and PSF-matched F150W reveals that the aperture photometry is robust when comparing \texttt{webbpsf} (pink) to individual bright ($<20.5$ ABmag) stars (greyscale).  In all cases, the CoGs agree at to $<$1\% for the adopted aperture diameter of 0.32 arcseconds (dashed vertical line). There exists an offset of $\sim$10\% for the F444W CoG within a 0.32 arcsecond diameter; the reader is cautioned to not use \texttt{webbpsf} for structural analyses with this data set. The median absolute deviation in the CoGs at a given radius for individuals stars is of order 0.5\% for F444W (top panel, bottom right) and 1\% for PSF-matched F150W (bottom panel, bottom right).}\label{fig:psfs}
\end{figure*}

We have used legacy field data to confirm that
empirical PSFs extracted from the latest reductions of the images are in excellent agreement with those generated through \texttt{webbpsf} (see also \citealt{Morishita23}, \citealt{Weibel24}). 
As an additional test, we compare the curve of growth (CoG) for \texttt{webbpsf} to those derived from moderately bright ($<$20.5 ABmag) individual stars selected across all fields.  The top panel of Figure~\ref{fig:psfs} shows results for F444W; we find that the \texttt{webbpsf} CoG is more concentrated than that for real stars, deviating by 10-20\% within a diameter of 0.32 arcseconds \citep[see also,][]{Weaver2024}.  However, deviations beyond this radius are less than 1\%, indicating that the photometry is reliable given the apertures adopted herein.  The field-to-field variation in the recovered CoG for individual stars is $<$1\%.  The bottom panel of Figure~\ref{fig:psfs} shows a similar test, but now considering the CoG for stars within the F150W PSF homogenized images.  Here, we see no significant offsets at any radius and a slightly higher field-to-field variation of order 1\%.  While these tests emphasize that our decision to use \texttt{webbpsf} does not impact the resulting photometry, we caution the reader for other applications.  Namely, \texttt{webbpsf} would not be adequate for a morphological analysis using this same data set as the underestimated PSF CoG in the cores would systematically bias the sizes larger. If it is not possible to generate an empirical PSF from real stars for these purposes, it is preferable to use an empirical PSF from other similar data sets \citep[e.g.,][]{Weaver2024}.

Finally, for ancillary HST filters we use empirical PSFs extracted from nearby legacy data using \texttt{EPSFBuilder} from the \texttt{photutils} python package \citep{Anderson2000, Anderson2016}.
Using the PSFs, we derive matching kernels using \texttt{pypher} \citep{Boucaud16} and convolve each image with the corresponding kernel.

The PSF-matched aperture fluxes are scaled to total fluxes in two steps: First, we scale them to the flux measured through a Kron-like aperture determined by \texttt{SourceExtractor} on the detection image and evaluated on a PSF-matched version of it and second, we divide them by the encircled energy of the Kron ellipse on the F444W PSF \citep[e.g.][]{Skelton14,Whitaker2019}.

To assess the depth of each pointing, we measure fluxes through 0.16\arcsec\ radius circular apertures in 5000 random positions without any nearby sources. The inferred 5$\sigma$ depths in each available filter are listed in Table \ref{tab:catalogs}. In this table we also provide the area of each association's footprint, which we assess based on the overlapping area between LW filters. By this metric, our survey obtained $\sim$432 sq arcmin of area with 6 or more filters of data for which we provide the photometric catalogs. 

We note that the LW overlapping footprint area can sometimes be smaller than the actual area of on sky data that is obtained, due to the prime-instrument's observing strategy and our program's optimization of pure parallel opportunities discussed in Section \ref{sec:realcycle1}.

Various flags are added to the catalogs. First, we flag all objects whose isophotal footprint as determined by \texttt{SourceExtractor} to overlap with one or more flagged pixels in the respective flag map. The latter are derived from the weight maps provided by \texttt{grizli}, flagging pixels with no weight. In addition, we grow the flag images through binary dilation in order to more generously flag objects near the edge of the mosaics or near regions with data quality issues.

We then flag stars using two different criteria. The first one identifies bright point sources (mag(F444W) $<$ 24.5) by requiring $1.2<{\rm F}(0.35\arcsec)/{\rm F}(0.16\arcsec)<1.4$ where ${\rm F}(0.35\arcsec)$ and ${\rm F}(0.16\arcsec)$ are the fluxes measured through circular apertures with radii of 0.16\arcsec\ and 0.35\arcsec\  respectively. To also flag fainter stars as well as diffraction spikes or contaminated sources around stars, we match our catalogs to sources with non-zero proper motion in the GAIA DR3 catalog \citep{GAIA2016, GAIA2023} using a search radius of 2.5\arcsec. 
Based on the \texttt{SourceExtractor} output, we flag sources with unreasonably large sizes (R$_{50}$(F444W) $>$ 50 pix) or Kron ellipses (circularized Kron radius $>$ 150 pix). This identifies spurious detections around bright stars as well as objects near stars or extended foreground sources contaminating their Kron ellipses. Further spurious detections and hot pixels are flagged using $0<{\rm R_{50}(F444W)}<1.2$ corresponding to unphysically small sizes. Finally, we flag objects that do not have a S/N $>$ 3 detection in \textit{any} of the available NIRCam wide filters, and objects with no valid photometric redshift fit (see next Section). All flags are combined into a \texttt{use\_phot} column, so that \texttt{use\_phot}=0, if any of the flags described above is set. In order for a source to have \texttt{use\_phot} $=1$, it is further required to not have any flagged pixels in any of the LW wide filters (F277W, F356W and F444W) that contribute to the stacked detection image. 
We use this latter criterion to estimate the survey area of each field, by counting the pixels which are not flagged in any of those 3 filters. The inferred survey areas are specified for each field in Table \ref{tab:catalogs}.

\begin{deluxetable*}{lllllllll}[!htbp]
\caption{5$\sigma$ depths (AB magnitudes) measured in circular apertures of 0.16\arcsec radius in each filter and the survey area in arcmin$^2$ of each pointing, for 6+ filter footprints covering novel sky area only.}\label{tab:catalogs}
\tablehead{\colhead{Field name}  & \colhead{F115W} &  \colhead{F150W}& \colhead{F200W} & \colhead{F277W} & \colhead{F356W}  & \colhead{F410M} & \colhead{F444W} & \colhead{Area$^{b}$ [arcmin$^2$]}  }
\startdata
j000352m1120 & 27.50 & 27.72 & 27.94 & 28.37 & 28.50 & - & 28.24 & 9.37 \\
j010408m5508 & 27.03 & 27.31 & 27.52 & 28.17 & 28.30 & - & 28.11 & 9.36 \\
j010500p0217 & 28.66 & 28.51 & 28.74 & 29.16 & 29.06 & 28.58 & - & 11.67 \\
j013444m1532 & 27.01 & 27.27 & 27.48 & 28.17 & 28.20 & - & 28.03 & 9.38 \\
j013748m2152 & 28.98 & 29.04 & 29.30 & 29.34 & 29.36 & 28.88 & 29.01 & 3.04 \\
j013900m1019 & 28.36 & 28.54 & 28.75 & 29.08 & 29.18 & - & 28.83 & 16.13 \\
j014428p1715 & 28.39 & 28.56 & 28.76 & 29.11 & 29.21 & - & 28.86 & 14.50 \\
j021716m0520 & 27.93 & 28.15 & 28.25 & 28.64 & 28.78 & - & 28.41 & 9.23 \\
j021728m0214 & 28.78 & 28.95 & 28.20 & 28.70 & 29.53 & - & 29.12 & 9.23 \\
j021800m0521 & 27.91 & 28.12 & 28.21 & 28.61 & 28.69 & - & 28.41 & 9.23 \\
j021824m0517 & 27.97 & 28.13 & 28.19 & 28.54 & 28.65 & - & 28.39 & 9.23 \\
j024000m0142 & 28.06 & 28.31 & 28.48 & 28.91 & 29.02 & - & 28.71 & 6.15 \\
j031404m1759 & 27.88 & 28.11 & 28.31 & 28.85 & 28.92 & - & 28.60 & 13.63 \\
j033212m2745 & 28.16 & 28.40 & 28.36 & 28.75 & 28.85 & - & 28.62 & 9.24 \\
j033224m2756 & 28.75 & 28.84 & 28.62 & 29.27 & 29.47 & 29.34 & 29.08 & 43.22 \\
j041616m2409 & 28.39 & 28.59 & 28.83 & 29.08 & 29.11 & - & 28.72 & 9.80 \\
j043844m6849 & 27.69 & 27.93 & 28.15 & 28.63 & 28.65 & - & 28.33 & 9.25 \\
j093144p0819 & 27.47 & 27.76 & 27.89 & 28.36 & 28.43 & - & 28.20 & 9.36 \\
j094232p0923 & 27.47 & 27.70 & 27.91 & 28.35 & 28.49 & - & 28.21 & 9.36 \\
j100024p0208 & 27.47 & 27.66 & 28.22 & 28.37 & 28.78 & - & 28.25 & 18.37 \\
j100040p0242 & 26.85 & 27.16 & 27.29 & 27.98 & 28.02 & - & 27.84 & 10.71 \\
j100736p2109 & 27.44 & 27.67 & 27.88 & 28.33 & 28.33 & - & 28.05 & 9.33 \\
j121932p0330 & 27.25 & 27.56 & 27.72 & 28.17 & 28.21 & - & 27.97 & 9.23 \\
j123140m1618 & 27.01 & 27.28 & 27.43 & 28.06 & 28.08 & - & 27.84 & 9.37 \\
j123732p6216$^{a}$ & 28.37 & 28.61 & 28.43 & 28.71 & - & 28.06 & 27.83 & 9.30 \\
j123744p6211 & 28.21 & 28.47 & 28.34 & 28.82 & 28.96 & - & 28.83 & 9.23 \\
j123816p6214 & 28.26 & 28.69 & 28.44 & 28.86 & 28.99 & - & 28.80 & 9.22 \\
j125652p5652 & 27.39 & 27.61 & 27.92 & 28.33 & 28.51 & - & 28.28 & 8.84 \\
j130016p1215 & 27.65 & 27.86 & 28.05 & 28.39 & 28.43 & - & 28.08 & 9.23 \\
j131432p2432 & 27.05 & 27.32 & 27.47 & 28.12 & 28.21 & - & 27.97 & 9.25 \\
j134348p5549 & 27.33 & 27.58 & 27.81 & 28.32 & 28.38 & - & 28.15 & 9.27 \\
j142536p5630 & 27.91 & 28.19 & 28.39 & 28.89 & 28.95 & - & 28.64 & 9.40 \\
j144904m1017 & 27.47 & 27.71 & 27.92 & 28.38 & 28.43 & - & 28.24 & 9.37 \\
j153500p2325 & 27.36 & 27.63 & 27.86 & 28.44 & 28.57 & - & 28.29 & 9.09 \\
j170720p5853 & 26.91 & 27.16 & 27.40 & 28.13 & 28.18 & - & 27.98 & 10.68 \\
j221648p0025 & 29.15 & 28.99 & 29.15 & 29.34 & 29.15 & 28.59 & 28.82 & 9.23 \\
j221700p0025 & 29.13 & 29.05 & 29.23 & 29.51 & 29.48 & 28.83 & 28.96 & 9.23 \\
j224604m0518 & 28.23 & 28.10 & 27.85 & 28.36 & 28.83 & - & 28.76 & 14.83 \\
j224616m0531 & 28.98 & 28.96 & 29.18 & 29.62 & 29.62 & 28.72 & 29.40 & 9.18 \\
j234512m4235 & 28.39 & 28.26 & 28.10 & 28.50 & 28.89 & 28.18 & 28.53 & 9.19 \\
\enddata
\tablecomments{  (a) filter depths listed correspond instead to F162M, F182M, F210M, F300M, F430M, F460M; see end of Section \ref{sec:reduction}. (b) Area is assessed based on the overlapping area of all LW filters, i.e. matches the area probed by our photometric catalogs. Actual area by total coverage of all filters is larger. We do not measure photometry from pointings where we obtained fewer than 6 filters, see Section \ref{sec:mishaps}. 4-filter pointings do not appear in this table.}
\end{deluxetable*}

\subsection{Photometric redshifts}\label{sec:eazy}

We run the template-based SED-fitting code \texttt{eazy} \citep{Brammer08} to estimate photometric redshifts for all the galaxies in our catalogs. To allow for some flexibility in the fitting and account for possible systematic uncertainties, we adopt an error floor of 5\% on all the fluxes. We use the \texttt{blue\_sfhz} template set\footnote{\url{https://github.com/gbrammer/eazy-photoz/tree/master/templates/sfhz}} which consists of 14 templates, 13 of which are generated through the Flexible Stellar Population Synthesis code \citep{Conroy2009, Conroy2010}. In addition, one template is derived from a recently obtained NIRSpec spectrum of an extreme emission line galaxy at $z=8.5$ \citep{Carnall23}. We allow the best-fitting redshift to vary in the range $z\in(0.01,20)$ and we run three iterations of the internal zeropoint optimization which corrects for possible calibration offsets based on the template fitting residuals. This yields small correction factors ranging from $\sim1$ to $\sim8$\% in all filters and pointings, except for F115W in j043844m6849 ($\sim12$\%) and F150W and F200W in j010500p0217 ($\sim11$\% and $\sim16$\%).

\subsubsection{Photometric redshift performance}\label{sec:photoz_performance}

In this section we simulate the photometric redshift performance of typical pointings in our survey. We characterize this using the JAdes extraGalactic Ultradeep Artificial Realizations \citep[JAGUAR;][]{Williams18}.
The individual realizations of JAGUAR mock catalogs are limited in area ($\sim$120 sq arcmin), significantly smaller than the PANORAMIC combined survey footprint. Due to the rapid decline with redshift of the assumed pre-JWST luminosity function in JAGUAR \citep{Oesch18}, our target populations become increasingly rare in single JAGUAR realizations. Thus, we join 10 realizations in order to enable a better characterization of the performance at higher redshifts.

Because of the heterogeneity of our full dataset, we base our performance on two typical subsets of our survey. The two categories include: pointings which have 6 broad-band filters, and pointings where we obtained a 7th filter (F410M). For each survey subset, we take the average of the limiting flux detection limits in each filter across all pointings in the category. For the 6-filter survey, these depths correspond to 5$\sigma$ detection limits in the shortwave bands of 27.5, 27.8, 27.9 ABmag (F115W, F150W, F200W) and long wave bands 28.4, 28.6 and 28.3 (F277W, F356W, F444W). For the 7-filter survey, we find 28.9, 28.8, 28.8 (F115W, F150W, F200W) and 29.3, 29.3, 29.0 and 28.7 (F277W, F356W, F410M, F444W). We select JAGUAR sources with brightnesses that correspond to S/N $>$ 10 in the F444W filter (given our sensitivity limits), and include photometric scatter and uncertainty in the mock photometry according to the average rms. We then measure photometric redshifts using EAZY with the same setup as described in Section \ref{sec:eazy}.

\begin{figure*}
    \begin{center}
    \includegraphics[width=1.2\textwidth,trim=100 0 0 30, clip]{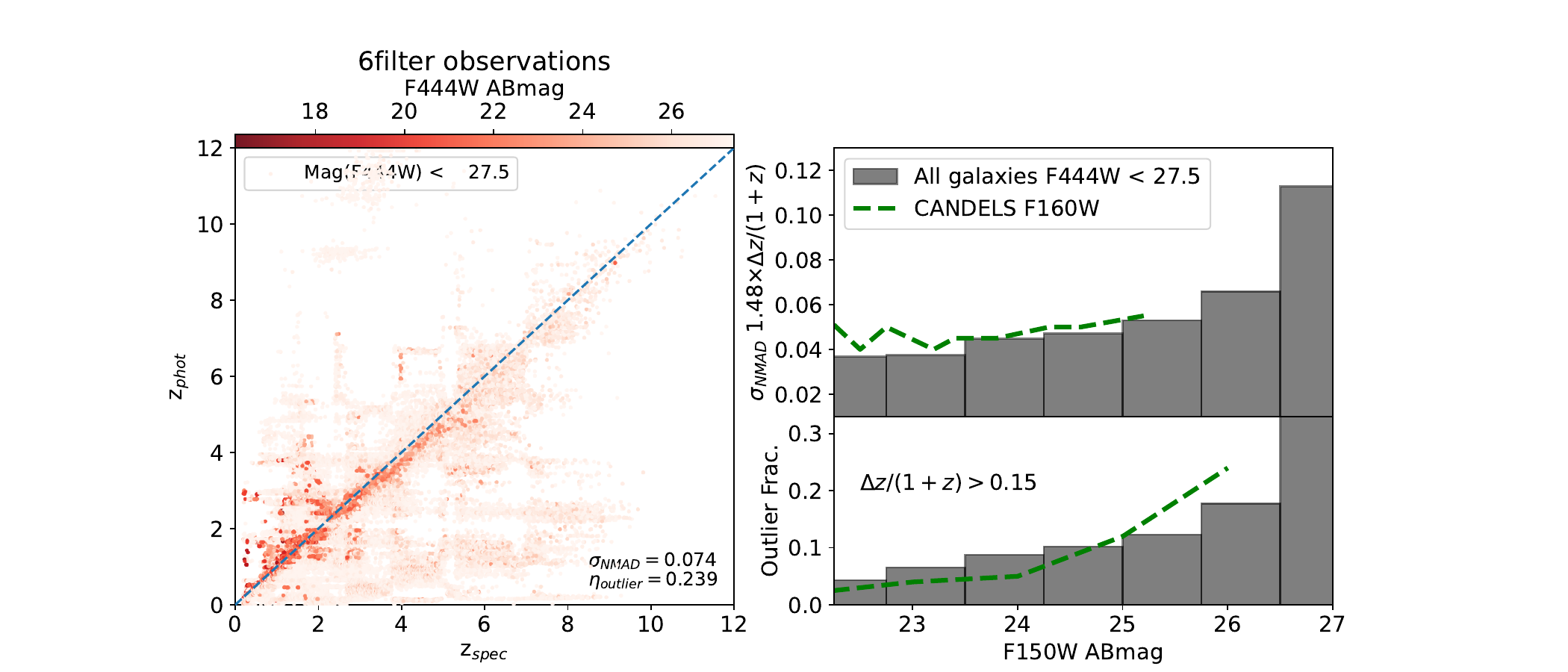}          \includegraphics[width=1.2\textwidth,trim=100 0 0 30, clip]{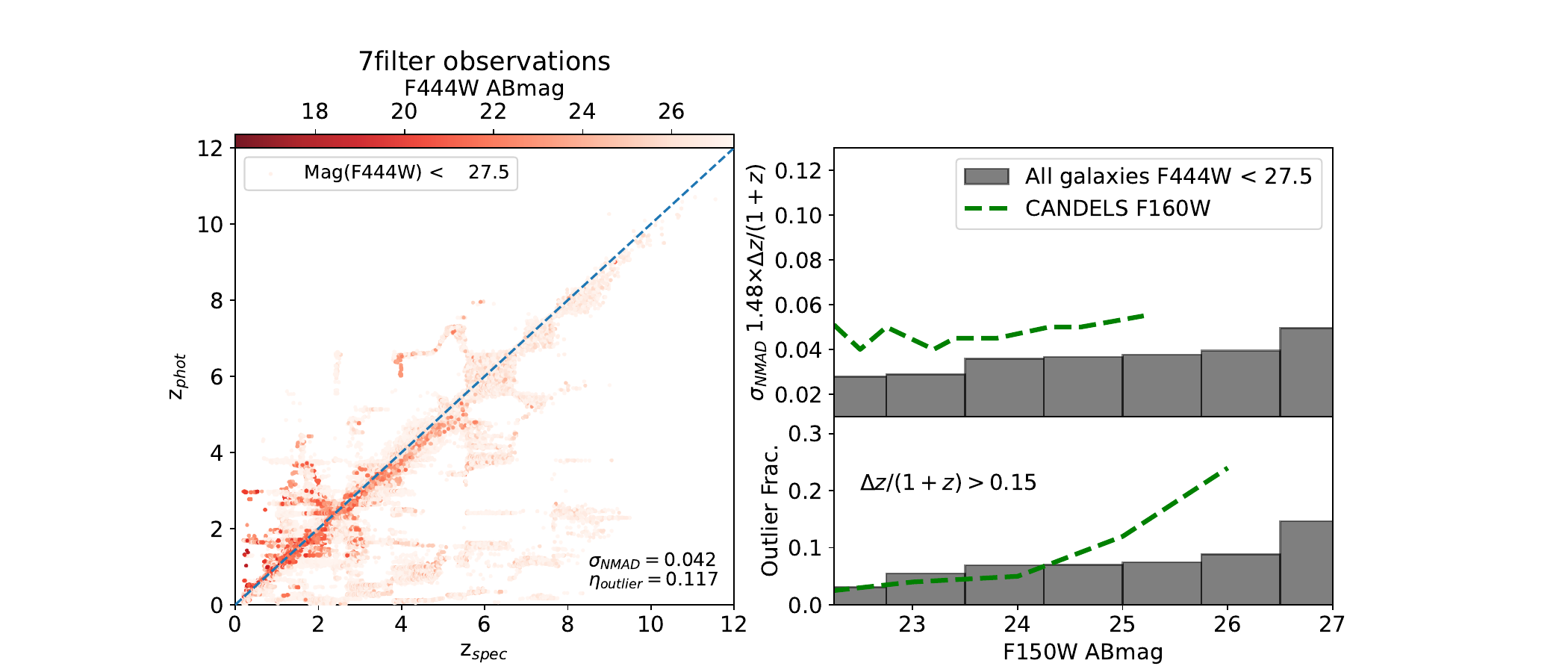}
    \caption{EAZY photo-z performance using JAGUAR simulation of our 6-filter (top row) and 7-filter strategy (bottom row). Left panels: points represent galaxies at various apparent magnitudes in F444W and are rank sorted by magnitude (brightest lie on top). Right panels: photo-z scatter ($\sigma_{NMAD}$) and outlier fraction (fraction of galaxies with $\sigma_{NMAD}>0.15$ as a function of F150W ABmag. Green curves show the typical scatter and outlier fraction as a function of HST H-band magnitude from CANDELS \citep{Dahlen13,Salvato19}, demonstrating that PANORAMIC achieves similar photometric redshift performance to HST legacy surveys without the need of ancillary data.}\label{fig:photoz_jaguar}
    \end{center}
\end{figure*}

In Figure \ref{fig:photoz_jaguar} we show the photometric redshift recovery for JAGUAR mock galaxies assuming our 6-filter baseline survey (top panels) and 7-filter (bottom panels). We characterize the photometric redshift performance using $\sigma_{\rm NMAD}$ defined as $\sigma_{\rm NMAD}=1.48\times\mathrm{median}\left(\frac{\vert \Delta z-\mathrm{median}(\Delta z)\vert}{1+z_\mathrm{spec}}\right)$
\label{eq:MAD} and outlier fraction defined as $f_\mathrm{outlier}=\Delta z / (1+z_{\rm{spec}})\ge 0.15$ 
\citep{Brammer08} where $z_\mathrm{spec}$ is the redshift of the mock galaxy and $\delta z$ is the difference between the photo-z and $z_\mathrm{spec}$. We find that our 6 (7)-filter survey has typical $\sigma_\mathrm{NMAD}=0.074$ (0.042) and typical $f_\mathrm{outlier}=0.24$ (0.12) (see Figure \ref{fig:photoz_jaguar}). We find that at F150W magnitudes $<$26 that were probed by the CANDELS survey \citep{Grogin11} our survey achieves similar $\sigma_\mathrm{NMAD}$ to that obtained by CANDELS \citep{Dahlen13, Bezanson2016, Salvato19}, even without the use of ancillary data. This clearly demonstrates the power of pure-parallel imaging in producing a wide-field infrared
survey that achieves CANDELS-like data quality but at longer wavelengths, opening a whole new realm of discovery space. 

Since the empirical galaxy evolution model that underlies the JAGUAR mock catalog is based on pre-JWST datasets, it inherently lacks populations of red galaxies across redshifts where pre-JWST datasets were less complete. This is particularly true for massive galaxies at $z>3$, which are primary science targets of this program (see Section \ref{sec:science}), although these are likely too rare to significantly impact the photometric redshift statistics. However, it is also the case that the lowest-mass  galaxies detectable by PANORAMIC at intermediate redshifts in JAGUAR have properties based on extrapolations of the stellar mass function from pre-JWST data, observations which were likely more incomplete at redder colors. Thus, we note the possibility that the low mass galaxies that dominate numbers across redshifts (and thus our photometric redshift statistics) are bluer with higher fraction detectable emission lines in JAGUAR than the real universe, making photometric redshifts appear easier to determine. 

\section{Science Goals} \label{sec:science}

The primary motivation for PANORAMIC are two-fold: to reveal statistical samples of (1) luminous galaxies at $z>9$, and (2) the most massive, red sources at $3<z<9$. Since the abundances of these populations are critical constraints on galaxy formation models, the pure parallel strategy has a strong advantage in lowering the uncertainties due to cosmic variance. As we will show in this section, this is true even though the data obtained as part of PANORAMIC is smaller in area than proposed. The area of PANORAMIC still increases the Cycle 1 area with $\geq$6 filters and F444W$<$28 ABmag by 60\%, with the key advantage that it does so using 40 independent sightlines.
As such, this dataset will enable a wealth of community science. 
Further, random sightlines enables new source discovery which is a key goal of PANORAMIC, since this is a first step before detailed followup can be launched for statistical samples of bright populations.
Here, we list only a few of the most important scientific questions that were motivating factors in the design of PANORAMIC.

\subsection{Unveiling the Brightest Galaxies at $9<z<12$}\label{sec:science_hiz}

\noindent While the first JWST observations have immediately pushed the cosmic frontier beyond HST's limit of $z\sim11$ \citep[e.g.][]{Oesch18}, our knowledge at these redshifts remains extremely uncertain. The number density evolution of galaxies at $z>10$ remains debated and is only based on a handful of spectroscopically confirmed galaxies to date (e.g. \citealt{CurtisLake23,Wang23a,Finkelstein2022, ArrabalHaro2023a, Castellano2024, Carniani2024}; for a review see \citealt{Adamo2024}). The star formation rate density (SFRD) at $z>10$ is found to be extremely high  based on early JWST-identified candidates at $z>10$ \citep[e.g.,][]{Naidu22, Atek23, Donnan22b, Harikane22}. These candidates lie more than an order of magnitude above the predicted luminosity functions  from simulations \citep[e.g., ][]{Finkelstein2024, Harikane2024}.
Interestingly, galaxy formation models largely diverge at $z>8$, resulting in predicted SFRDs that differ by factors of up to 50$\times$ by $z\sim12$ \citep[][]{Feldmann2024}. These divergences could reflect our uncertainty in the efficiency of galaxy growth in dark matter halos, 
or, could highlight a critical change in our understanding of early galaxy formation,  
including a lack of dust attenuation \citep[e.g.][]{ferrara2023, mauerhofer2023}, an evolving initial mass function \citep[e.g.][although see \citealt{Cueto2024}]{yung2024, trinca2024}, bursty star formation \citep[e.g.][]{mason2023,mirocha2023}, black hole contribution \citep{pacucci2022} and/or low-redshift interlopers \citep{ArrabalHaro23}. The extremely small sample sizes still prevent firm conclusions. Thus, measuring the UV luminosity function at $z>10$ with $\sim$2\micron\ imaging is a powerful test of these models -- in particular, when probing the most luminous sources \citep{Behroozi18}.

\begin{figure}[!t]
    \centering
    \includegraphics[scale=.45]{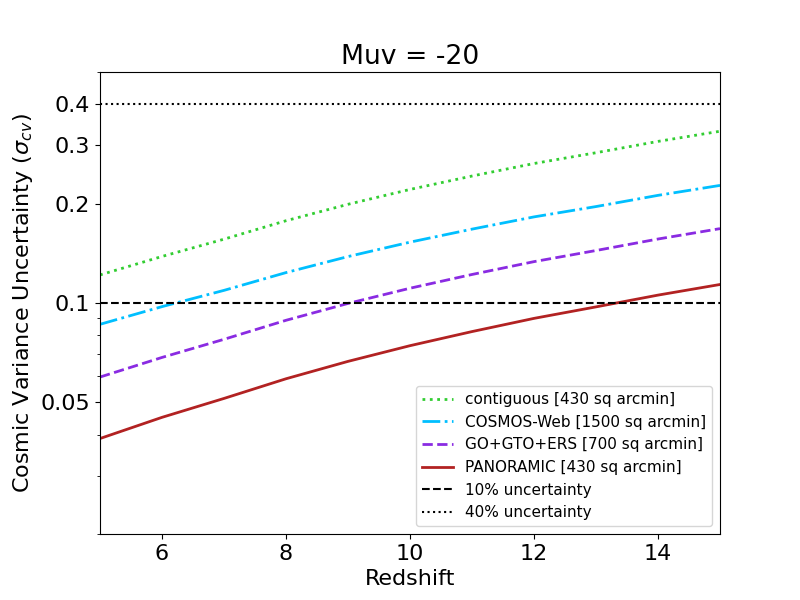}
    \caption{Cosmic variance uncertainty for galaxies with M$_{UV}$=-20 with redshift \citep{Trapp2020}, estimated for the 40 independent sightlines with PANORAMIC (red) compared to an equivalent contiguous survey of the same area, and to existing extragalactic surveys. }
    \label{fig:cv}
\end{figure}

PANORAMIC was designed to build a wide-area extragalactic tier capable of identifying statistical samples of bright, $z>9$ galaxies that make meaningful improvements to abundances and population-level census uncertainties, while providing spectroscopic targets that are bright enough to contribute significant astrophysical constraints on galaxy formation.  
In our original survey design, we expected to identify $\gtrsim$12  galaxies of similar brightness to GNz-11,  
and up to 80 new $z\geq9$ galaxies at $\gtrsim\!L^*$, as tracers of early reionization. 
Due to their highly biased nature, the most luminous sources are subject to the largest cosmic variance uncertainties, which limits the ability of smaller fields to constrain models. 
However, we find that our survey will still put meaningful constraints even at $\sim$1/3 of the proposed size. 
To estimate the cosmic variance uncertainty ($\sigma_{\textrm{cv}}$) of PANORAMIC we use the \texttt{galcv} code from \citet{Trapp2020} using our real association areas from Table \ref{tab:catalogs}. We combine the individual cosmic variances of each of our pointing following \citet{Moster2011} for separate, and uncorrelated fields, which with N=40 pointings roughly improve the uncertainty compared to each individual pointing as $\sqrt{N}$. Figure \ref{fig:cv} shows the excess uncertainty due to cosmic variance in our survey is a factor of 3 improvement at $z\sim12$ than a comparable survey of contiguous area. We additionally include estimates for the existing GO+GTO+ERS and COSMOS survey. 
This plot demonstrates that PANORAMIC will make meaningful improvements to abundance, census, and luminosity function estimates.

In Section \ref{sec:design}, we demonstrated that  parallel NIRCam imaging of the scale originally proposed for PANORAMIC ($\sim$0.4 sq degrees using 3 filter-sets) 
are feasible within one JWST cycle. This indicates that the science originally proposed will soon be possible using accumulated parallel imaging on single-cycle timescales in the future.

\subsection{ The census of red massive galaxies at $z>3$}\label{sec:science_redgal}

Because HST was only sensitive to the rest-frame UV light of $z>3$ galaxies, the pre-JWST census of galaxies in the first 2 Gyr of cosmic history remained highly biased to relatively unobscured, star-forming galaxies. 
Additionally, the discovery of significant numbers of (more massive) IRAC- or ALMA-selected galaxies over large areas that remained undetected with HST demonstrates a striking, and still little explored discovery space \citep{Caputi12, Caputi15, Williams19, Wang19, Manning22}. Pre-JWST surveys were largely incomplete at the extremely red rest-optical colors of these $z>3$ massive galaxies.

The first small-area JWST images immediately revealed very red galaxies missed by HST, including dust obscured and quiescent candidates at $z\sim3-9$. The number densities, typical masses, and redshift distributions of these red sources are still highly uncertain \citep[e.g.][]{McKinney2024}, but support pre-JWST conclusions 
that HST surveys might have missed up to 90\% of  massive galaxies \citep[log$_{10}$M$_*$/M$_{\odot}>10.5$ at $z>3$;][]{Wang19} because the massive end is almost entirely optically faint \citep{Weibel24, Gottumukkala23, Harvey2024}. Surprisingly, JWST-selected red sources span lower than expected stellar masses \citep[$9<$ log$_{10}$M$_*$/M$_{\odot}<10.5$;][]{Barrufet22, Barrufet24, Rodighiero23, PerezGonzalez22, Williams2023b} based on dust scaling relations from lower redshift \citep{Whitaker2017}. 
These discoveries indicate that the incompleteness in HST selection clearly does not only impact the high-mass end, and highlights the importance of wedding cake survey strategies (e.g. PANORAMIC along with other surveys, see Figure \ref{fig:areadepth}).

Despite the short cosmic time available for dust production at high-redshift, early galaxies host surprisingly abundant dust \citep{Fudamoto21, Schouws2022,Xiao23,McKinney2023,Akins23}. This explains why even the deepest HST+Spitzer campaigns missed these ``dark'' galaxies, but brings new questions to the forefront. Some fraction of newly-discovered red galaxies at $3<z<9$ show higher than expected efficiency of baryon conversion into stars \citep{Labbe23a, Xiao23}, while exhibiting dramatic field-to-field variation \citep{Desprez2024}. Others exhibit surprisingly dominant AGN signatures whose black holes are more massive than expected, appearing as Little Red Dots \citep[LRDs;][]{Matthee23,Labbe2023b,Greene24, Furtak23b}. However, a gap exists between the lower-mass tail of the distribution probed by JWST deep-fields and the $\gtrsim$degree scale areas tracing the most extreme quasars \citep{Kokorev2024, Inayoshi2024}.  In particular at $z\sim7$ we totally lack constraints on the abundances and luminosities at  L$_\mathrm{bol}\sim 10^{46}-10^{47} \textrm{L}_{\odot}$  scales (between LRDs and massive QSOs), and PANORAMIC provides the opportunity to discover such candidates, the first step to  quantifying the strain on both black hole seeding and early growth models \citep{Inayoshi20} while mitigating cosmic variance to enable robust conclusions.

Collectively,  these surprises require larger sample sizes to adequately interpret the implication for our theories about black hole formation and galaxy growth. 
PANORAMIC stands to address these important issues:  $\geq$6 broadband filters have proved remarkably efficient at identifying LRDs, massive dusty galaxies, and quiescent galaxies at $z>3$ without the need for ancillary data \citep[e.g.][]{Greene24, Long2023, Williams2023b}. 
For all classes of red galaxies at this poorly sampled epoch, probing larger volumes to collect larger and brighter samples, while using increased numbers of sight lines to decrease abundance uncertainties, is the best way to improve the empirical picture.

\begin{figure*}[!t]
    \centering
    \includegraphics[width=1\textwidth]{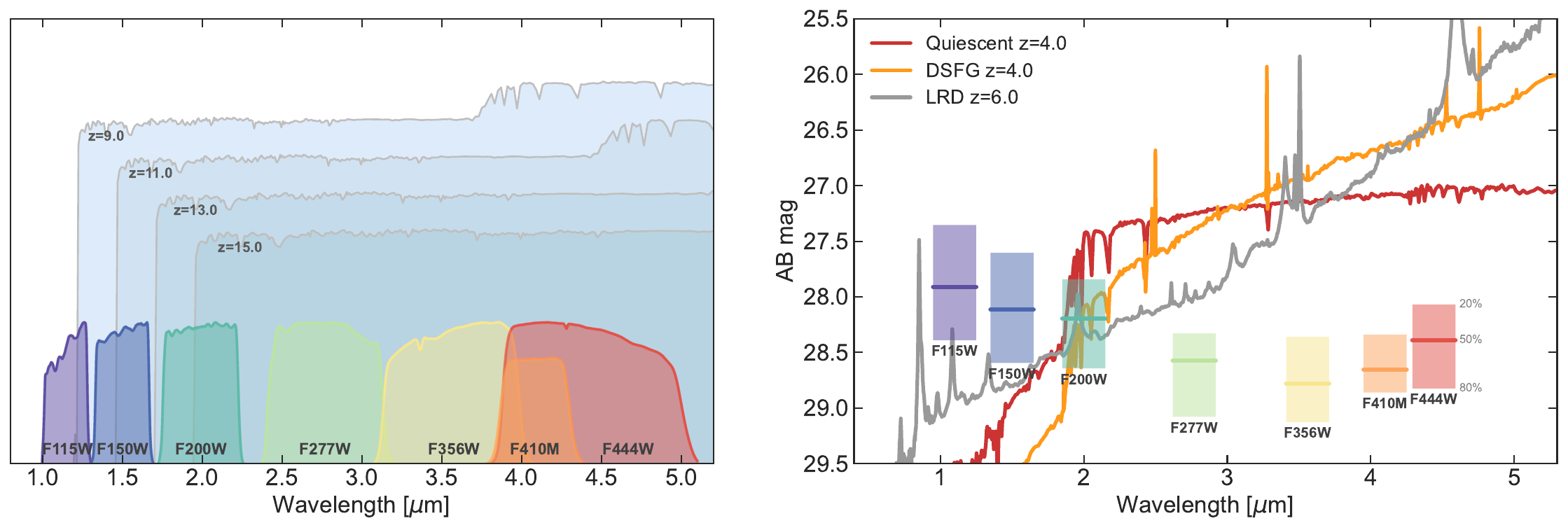}
    \caption{Target galaxy SEDs and PANORAMIC survey design. \textit{Left:} High redshift Lyman break galaxies together with the filter set of PANORAMIC. The strong spectral break at the redshifted Ly$\alpha$ line allows us to identify $z\gtrsim8$ galaxies as dropouts in successively redder filters. \textit{Right:} The sensitivity limits of PANORAMIC are shown together with typical galaxy types of interest at $z\gtrsim3$: quiescent galaxies (red), dusty star-forming galaxies (orange) and so-called little red dot AGN candidates (gray). The vertical bars extend from the 20\% deepest to the 20\% brightest fields of the final PANORAMIC survey, with the center line denoting the median 5$\sigma$ depth (as measured in fixed circular apertures of 0\farcs16 radius in empty sky regions).  }
    \label{fig:SEDs}
\end{figure*}

A key feature of the design of PANORAMIC is the capability to distinguish  various red galaxy populations at $3<z<9$. In Figure \ref{fig:SEDs} (right panel) we show a range of example SEDs with our average imaging depths (see Section \ref{sec:photoz_performance}) demonstrating that the data have the capability to distinguish the SED shapes, where existing wide-area ground-based broad band NIR and Spitzer imaging were largely unable to do so without spectroscopy \citep[e.g.][]{Forrest20, Marsan2022, Glazebrook17, Nanayakkara2024}. Conveniently the optimal selection for high-redshift sources (science targets discussed in Section \ref{sec:science_hiz}) coincides with the optimal design for red galaxies at $3<z<9$ discussed here, given the need for deep NIR coverage where the galaxies are faint, and well-sampled SEDs between 1-5$\mu$m to distinguish blue from red continuum in the longer wavelength bands.

\subsection{Discovery Space}

Pursuing novel sky area with pure parallel imaging presents an exciting opportunity to discover unknown unknowns, and open up poorly explored parameter space of known populations of objects. Here we list a few categories of such objects we have identified using PANORAMIC beyond those discussed in the previous sub-sections, to illustrate the immense potential of, and motivate the collection of additional areas with pure parallels.

\subsubsection{Intrinsically bright sources at $z>14$}

The first JWST imaging datasets revealed possible ultra-luminous galaxy candidates at $z>14$ \citep[e.g.,][]{Donnan22b, Naidu22} that are seemingly inconsistent with $\Lambda$CDM. Although one candidate was ultimately an improbable $z\sim5$ contaminant \citep{ArrabalHaro23}, more successful candidates continue to be identified at magnitudes accessible by PANORAMIC \citep[e.g. $z_{spec}=14.3$ and F444W $<$ 27.3 ABmag;][]{Carniani2024}. Continuing the search over wide area will, critically, improve the existing limits on abundances even at $z\sim14$ \citep[e.g.][]{Robertson2024}. 
Our filter combination will allow robust 3-filter selection of dropout sources to $z=30$ (in principle), enabling a fast characterization of abundances at these early epochs. Even upper limits from non-detections over wider areas, with improved uncertainties from cosmic variance, provide valuable baselines for interpreting the results from deeper narrow surveys.

\begin{figure*}
\centering
    \includegraphics[width=0.9\textwidth]{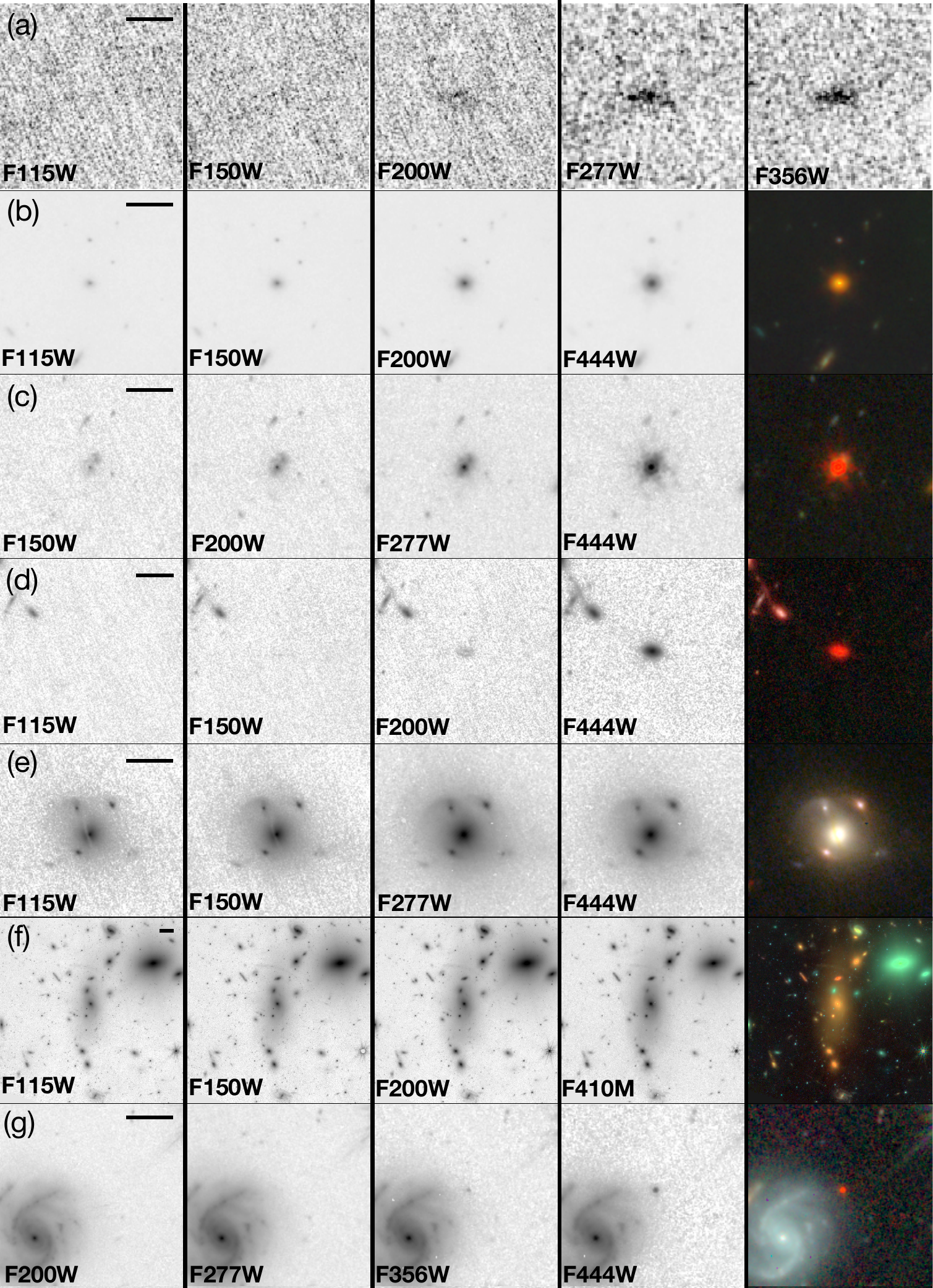}
    \caption{  Exceptional sources discovered from parallel imaging highlighting importance of wide-areas. Top to bottom: (a) candidate F150W-dropout; (b) candidate quiescent galaxy at $z>3$; (c) intrinsically bright (F444W$\sim$20 ABmag) LRD; (d) The reddest (lower limit to color F150W-F444W $>$ 6.4 ABmag) galaxy at $z\sim3.4$ in our survey; its color exceeds than of any other source in the JADES, PRIMER and CEERS areas combined ($\sim$730 square arcmin); (e) $z=3.9$ twice-lensed galaxy-galaxy lens; (f) a z$\sim$1 galaxy cluster candidate with newly revealed lensing arcs; (g) an F444W$\sim$24 ABmag Y-dwarf candidate. Horizontal scale bar in each row indicates 1\arcsec. }
    \label{fig:coolstuff}
\end{figure*}

\subsubsection{Discovery of strong lensing structures}

Probing wider areas increases the chances for identifying other rare phenomena typically relegated to ground-based all-sky (or square degree-scale) surveys. PANORAMIC has identified 4 strong galaxy-galaxy lens candidates with sources likely at $z>3$ (e.g. Figure \ref{fig:coolstuff}, panel e). Additionally, one pointing by chance overlaps a candidate $z\sim1$ galaxy cluster in existing DECam Legacy Survey DR9 imaging \citep[][]{Dey2019, Zhou2021, Zhou2023}, but the new near-IR imaging from PANORAMIC now reveals multiple lensing arcs indicating a likely galaxy cluster (Figure \ref{fig:coolstuff}, panel f).

\subsubsection{Unknown unknowns}

Wide parallel imaging  opens the opportunity to explore the fringes of the distribution of physical properties of known types of sources. For example, among our data we identified the reddest galaxy in the known universe (panel d; Figure \ref{fig:coolstuff}) which, thanks to our deep upper limits in F150W, puts a lower limit to the F150W-F444W color at 6.35, despite its very bright F444W magnitude (22 ABmag). While such red colors are a bit more common at fainter magnitudes \citep[F444W$>$25, e.g.][]{Williams2023b}, no similar source (to our knowledge) exists in similar extragalactic data. 

Targeting new, unexplored areas of the sky (compared to revisiting regions that already have 3-5$\mu$m imaging from Spitzer) opens up the potential for new unexpected discoveries. Numerous PANORAMIC pointings contain unexpected surprises. As examples, PANORAMIC has identified several 3-micron dropouts (F356W-F444W $>$ 1.6 and F444W $<$ 24-26.5; Figure \ref{fig:coolstuff} panel g). While these are more plausibly cool stars  which have similarly red F356W-F444W $>$ 1.6 colors \citep[e.g. Y-dwarfs; ][]{Beiler2023, Beiler2024}  than $z\sim$30 Lyman-break dropouts, given their extremely bright magnitudes, to our knowledge such objects are uncommon among the deep field datasets. This may be because deep fields typically point outside of the plane of the Milky Way, only probing halo stellar populations. In any case, any of the plausible interpretations of F356W-dropouts (bright halo brown dwarf, very red-continuum strong emission line galaxies, z$\sim$30 candidates) are likely to be rare occurrences among deep-field datasets, making these identifications and their followup interesting for a variety of astronomical sub-fields.

This potential should motivate further ambitious programs to collect wide-area imaging for random sightlines.

\section{Summary \& Data Release} \label{sec:DR}

We have presented and characterized the PANORAMIC Survey, the first JWST NIRCam extragalactic pure parallel imaging program in Cycle 1. This program obtained $> 430$ sq arcmin of imaging in novel sky area,  increasing the total extragalactic area observed in Cycle 1 by 60\% (compared to the total area of all Cycle 1 programs including GO, GTO and ERS with 6+ filters). The data are deep, with excellent photometric redshift performance, demonstrating the power of the pure parallel observing mode for collecting wide and deep, scientifically informative, imaging data. This is all possible at negligible cost to observatory resources, while representing the largest science-time investment of any extragalactic GO imaging program (including those with fewer filters) by more than a factor of 2 ($\sim190$ hours of science time on sky).

We also characterize the properties of this survey using our first data release (DR1), which includes the following science-ready data products. For data which probes novel sky area using JWST/NIRCam, we release mosaics in each filter. For images that entirely overlap other JWST/NIRCam surveys, our data will appear in legacy mosaics released in the DJA (with exception of COSMOS-Web footprints, where our data represents the longest integration time on sky, we present co-added mosaics). We also generate fitsmap viewers for each footprint \citep{HausenRobertson2022}.

The mosaics and fitsmaps are currently available at \url{https://panoramic-jwst.github.io/}. Upon acceptance of this manuscript our refereed data products, including photometric catalogs, will be hosted on MAST as high level science products (HLSP) at \url{https://doi.org/10.17909/fpzr-as35}

The PANORAMIC survey demonstrates the unique power of pure parallel imaging programs, which can be an extremely efficient and important tool to obtain a wide-area legacy imaging dataset with JWST for a wealth of science and follow-up studies by the community, from the local Universe (e.g., the environment around the LMC) up to the most distant galaxies.

\section*{Acknowledgments}
The PANORAMIC team gratefully acknowledges Shelly Meyett at STScI who's efforts enabled the science pure parallels mode on JWST, only operational for the first time in Cycle 1 to be successful. Blair Porterfield is thanked for program support.

This data release and science-ready data products makes use of archival JWST imaging from the following programs: 1180, 1181, 3215 (PI Eisenstein); 1210 (PI Luetzgendorf); 1283 (PI Oestlin); 1727 (PI Kartaltepe); 1810 (PI Belli); 1837 (PI Dunlop); 1895 (PI Oesch); 1963 (PI Williams); 2198 (PI Barrufet); 3383 (PI Glazebrook); 6541 (PI Egami).  The authors acknowledge the following teams and PIs for developing their observing program with a zero-exclusive-access period: COSMOS-Web (PIs Jeyhan Kartaltepe \& Caitlin Casey), PRIMER (PI James Dunlop), JEMS (PIs Christina Williams, Michael Maseda, Sandro Tacchella), FRESCO (PI Pascal Oesch), CEERS (PI Steve Finkelstein), JOF (PIs Daniel Eisenstein \& Roberto Maiolino), PID 6541 (PI Eiichi Egami). This work is based in part on observations made with the NASA/ESA/CSA James Webb Space Telescope.

This research is based on observations made with the NASA/ESA Hubble Space Telescope obtained from the Space Telescope Science Institute, which is operated by the Association of Universities for Research in Astronomy, Inc., under NASA contract NAS 5–26555. This data release and science-ready data products makes use of archival HST imaging from programs with PID 10092, 10189, 10258, 10339, 10340, 10530, 11108, 11144, 11359, 11600, 12007, 12060, 12061, 12062, 12064, 12099, 12177, 12190, 12328, 12385, 12440, 12442, 12443, 12444, 12445, 12459, 12461, 12578, 12866, 12990, 13002, 13063, 13294, 13386, 13420, 13459, 13496, 13641, 13657, 13779, 13790, 13792, 13793, 13868, 13871, 13872, 14038, 14043, 14114, 14209, 14216, 14496, 14721, 14747, 14895, 15115, 15117, 15229, 15363, 15647, 15663, 15862, 15936, 16259, 16278, 16872, 9352, 9425, 9480, 9500, 9583, 9584, 9727, 9728, 9803, and   9822.

This research was supported by the International Space Science Institute (ISSI) in Bern, through ISSI International Team project 562 (First Light at Cosmic Dawn: Exploiting the James Webb Space Telescope Revolution). Ideas for this project were conceived in part at meetings made possible by the Sexten Center for Astrophysics and the Aspen Center for Physics. 

This work is based in part on observations made with the NASA/ESA/CSA James Webb Space Telescope. The data were obtained from the Mikulski Archive for Space Telescopes at the Space Telescope Science Institute, which is operated by the Association of Universities for Research in Astronomy, Inc., under NASA contract NAS 5-03127 for JWST. These observations are associated with program 2514. Support for program JWST-GO-2514 was provided by NASA through a grant from the Space Telescope Science Institute, which is operated by the Association of Universities for Research in Astronomy, Inc., under NASA contract NAS 5-03127.  

The work of CCW is supported by NOIRLab, which is managed by the Association of Universities for Research in Astronomy (AURA) under a cooperative agreement with the National Science Foundation.  
This work has received funding from the Swiss State Secretariat for Education, Research and Innovation (SERI) under contract number MB22.00072, as well as from the Swiss National Science Foundation (SNSF) through project grant 200020\_207349. The Cosmic Dawn Center (DAWN) is funded by the Danish National Research Foundation under grant DNRF140. PD acknowledges support from the NWO grant 016.VIDI.189.162 (``ODIN") and warmly thanks the European Commission's and University of Groningen's CO-FUND Rosalind Franklin program. AH acknowledges support by the VILLUM FONDEN under grant 37459. The Cosmic Dawn Center (DAWN) is funded by the Danish National Research Foundation under grant DNRF140. 
MVM acknowledges support from the National Science Foundation via AAG grant 2205519 and the Wisconsin Alumni Research Foundation via grant MSN251397.

This paper uses data that were obtained by The Legacy Surveys: the Dark Energy Camera Legacy Survey (DECaLS; NOAO Proposal ID \# 2014B-0404; PIs: David Schlegel and Arjun Dey), the Beijing-Arizona Sky Survey (BASS; NOAO Proposal ID \# 2015A-0801; PIs: Zhou Xu and Xiaohui Fan), and the Mayall z-band Legacy Survey (MzLS; NOAO Proposal ID \# 2016A-0453; PI: Arjun Dey). DECaLS, BASS and MzLS together include data obtained, respectively, at the Blanco telescope, Cerro Tololo Inter-American Observatory, National Optical Astronomy Observatory (NOAO); the Bok telescope, Steward Observatory, University of Arizona; and the Mayall telescope, Kitt Peak National Observatory, NOAO. NOAO is operated by the Association of Universities for Research in Astronomy (AURA) under a cooperative agreement with the National Science Foundation. Please see http://legacysurvey.org for details regarding the Legacy Surveys. BASS is a key project of the Telescope Access Program (TAP), which has been funded by the National Astronomical Observatories of China, the Chinese Academy of Sciences (the Strategic Priority Research Program "The Emergence of Cosmological Structures" Grant No. XDB09000000), and the Special Fund for Astronomy from the Ministry of Finance. The BASS is also supported by the External Cooperation Program of Chinese Academy of Sciences (Grant No. 114A11KYSB20160057) and Chinese National Natural Science Foundation (Grant No. 11433005). The Legacy Surveys imaging of the DESI footprint is supported by the Director, Office of Science, Office of High Energy Physics of the U.S. Department of Energy under Contract No. DE-AC02-05CH1123, and by the National Energy Research Scientific Computing Center, a DOE Office of Science User Facility under the same contract; and by the U.S. National Science Foundation, Division of Astronomical Sciences under Contract No.AST-0950945 to the National Optical Astronomy Observatory.

\appendix

\section{Known issues}\label{sec:mishaps}
 
Beyond the pressure on the DSN, the implementation of science pure parallels presented a few additional early challenges to parallel data collection. While these challenges ultimately were resolved quickly, here we briefly track the impacted parallel observations. Observations that were skipped or rescheduled are marked as withdrawn in APT and the visit status page\footnote{\url{https://www.stsci.edu/cgi-bin/get-visit-status?id=2514&markupFormat=html&observatory=JWST}}.

\subsection{Skipped observations}

A number of single exposures were skipped for various reasons, but for which we still obtained data in at least one exposure per filter. These include the eastern most pointing in association j033212m2745 which lost exposures in F200W and F444W when the prime visit (PID 1211, visit 1, observation 16) lost lock on the guide star. 
In total, we have 4 visits with only 1 exposure, corresponding to our program's observation numbers 264, 265, 112, 076. 

\subsection{NIRSpec shorts}

A high fraction of our data was taken in parallel with NIRSpec multi-shutter array (MSA) observations. MSA electrical shorts are known to corrupt NIRCam imaging obtained in parallel, and several of our NIRCam exposures were impacted by this effect, with an especially high fraction in GOODS-North. 

\subsection{Guide star failures}

For a few pointings, a guide star failure for the prime program forced rescheduling of the observations. For prime-program observations based on MSA \citep[e.g. the NIRSpec GTO Wide program;][]{Maseda2024} these observations were repeated at the same orientation angle. However, in at least one case (j013748m2152) this resulted in the repeated observations being taken at a different observatory position angle, causing 2 of the filter footprints to be significantly rotated relative to the other filters, decreasing the area covered by all filters simultaneously. 

\subsection{Artifacts}

Several pointings exhibit scattered light features \citep[e.g. wisps, claws, dragons breath;][]{Rigby2023} that were not adequately or effectively removed using existing scattered light templates. In the case of j041616m2409 (a field also targeted by the CANUCS survey) the module and pointing most impacted sits on top of a CANUCS pointing that was uncorrupted. 
For simplicity we have removed this module's data for building our catalogs in the DR1 data release since uncorrupted data over the entire module exists elsewhere. 

We also note that pointing j130016p1215 is affected by a persistence image of Saturn from an earlier observation. We did not attempt to remove persistence artifacts in this data release.

\subsection{Unforeseen complications} 

A series of updates to the scheduling software and the on-board operating system in March 2023 resulted in skipped parallel observations (corresponding to 2 filters of parallel imaging) per prime-instrument visit until a solution was identified a few weeks later. This impacted $\sim$20 of our parallel pointings, which as a result were observed only in 4 filters instead of 6. This primarily impacted our pointings near GOODS-N since this time period spanned the scheduled observations for the NIRSpec GTO-Wide program in this field \citep{Maseda2024}. Four pointings in novel sky area were also impacted and thus only have 4 filters: j145652p2444, j090000p0207, j094232p0923, j150604p5409.

During this timeframe, an update to the operating system on board the spacecraft also resulted in 5 of the parallel pointings from prime program PID 1213 (PANORAMIC program observation numbers 23-25 and 35-46) getting skipped entirely in March 2023.

\bibliography{manu}{}
\bibliographystyle{aasjournal}

\end{document}